\documentclass[10pt, journal]{IEEEtran}

\usepackage{epsfig}
\usepackage{amsmath}
\usepackage[T1]{fontenc}
\usepackage{times}
\usepackage{amsmath,bm}
\usepackage{array}
\usepackage{tabularx}
\usepackage{makecell}
\usepackage{comment}
\usepackage{multirow}

\usepackage{caption}
\usepackage{subcaption}
\IEEEoverridecommandlockouts

\fontfamily{ptm}\selectfont

\begin{document}

\title{On the Fairness of Wi-Fi and LTE-LAA Coexistence}


\author{\IEEEauthorblockN{Morteza Mehrnoush, Sumit Roy}\\
\IEEEauthorblockA{Department of Electrical Engineering, University of Washington, Seattle, WA 98195, USA\\
Email: \{mortezam, sroy\}@uw.edu} \thanks{This work was supported by the National Science Foundation (NSF) under grant 1617153.}}

\maketitle


\begin{abstract}
With both small-cell LTE and 802.11 networks now available as alternatives for deployment in unlicensed bands at $5$ GHz, investigation into their coexistence is a topic of great interest. 3GPP Rel. 14 has standardized LTE licensed assisted access (LAA) that seeks to make LTE more coexistence friendly with Wi-Fi by incorporating listen before talk (LBT). However, the {\em fairness} of Wi-Fi and LTE-LAA sharing is a topic that has not been adequately explored. In this work, we first investigate the 3GPP definition of fair coexistence in \cite{3GPP_TR} via new analytical models. By tuning the LTE-LAA parameters, we exemplify scenarios when the 3GPP notion of fairness is achieved and conversely, when not achieved. The formal notions of access and proportional fairness is then considered for these scenarios to compare and contrast with the 3GPP definition. 

\end{abstract}



\section{Introduction}

Provisioning of high bandwidth end-user access via co-located small cell LTE or 802.11 WLAN networks has spurred great interest as to how they may (time) share the spectrum in the 5 GHz UNII bands where nearly 600 MHz is now identified for unlicensed use \cite{sharing, FCC}. Wi-Fi networks have been architected from inception for sharing among in-network nodes via the carrier sense multiple access collision avoidance (CSMA/CA) - a distributed random access mechanism. In contrast, LTE was originally designed for high efficiency licensed operation based on a centralized scheduler that allocates network resources among in-network nodes. While Wi-Fi's CSMA/CA mechanism naturally extends to spectrum sharing with other networks, new specifications were required to achieve sharing by LTE, since the latter was not originally designed to share with non-LTE systems. 

There are currently two specifications for unlicensed LTE operation: a) LTE Unlicensed duty cycling (LTE-U DC) and b) LTE Licensed Assisted Access (LTE-LAA). LTE-U employs a duty-cycle based approach along with Carrier Sense Adaptive Transmission (CSAT) to adapt to the LTE-U duty cycle according to the WLAN load \cite{LTEU}. LTE-LAA was standardized by 3GPP that integrates Listen-Before-Talk (LBT) mechanism \cite{3GPP_TR, ETSILAA17} - making it similar to CSMA for WLANs - to enable spectrum sharing worldwide in markets where it is mandated. LTE-U, on the other hand, is proposed for regions where LBT is not required and is promoted by LTE-U forum \cite{LTEU}, an industry SIG and is not a formal standard. As currently specified, LTE-LAA and LTE-U intend to utilize LTE carrier aggregation feature for enhanced data throughput on both  downlink and uplink for LTE-LAA and downlink only for LTE-U. 

A fundamental aspect of coexistence among {\em dissimilar} networks such as Wi-Fi and LTE-LAA is whether the sharing is {\em fair} in any acceptable sense. Clearly, there are several well-accepted notions of sharing among networks - among which min-max and proportional fairness \cite{WLAN_Fairness2005, WLAN_Fairness2008, Fairness_11e} are well-recognized. The 3GPP definition \cite{3GPP_TR} states ``LAA design should target fair coexistence with existing Wi-Fi networks to not impact Wi-Fi services more than an additional Wi-Fi network on the same carrier, with respect to throughput and latency''. In other words, the definition merely imposes a {\em insensitivity} requirement on Wi-Fi performance and is oblivious to actual LTE-LAA network throughput achieved in such sharing scenarios. As we will show, this does not lead to ``fair sharing'' under many circumstances. There are {\em numerous} factors on {\em both} sides that impact any reasonable approach to fairness, and an adequate definition is needed that recognizes this complexity and seeks to balance the rights of two rather dissimilar networks. 

The 3GPP definition (which is quite different from traditional notions of fairness) has led to considerable dissonance among the industrial research community. Given a fairness definition consecrated in the standard, it automatically becomes a target to be met; however the inherent definitional shortcomings like those noted above are an impediment to a more meaningful discussion of ``true fairness'' and how it may be achieved\footnote{As a result, claims by industry players appear to be tinged with partisan interests, i.e. promoting one side over the other, using hand-picked examples.}. Our work hope is to fill this important void in a meaningful manner, i.e. by bringing a fundamentally unbiased perspective that provides meaningful inputs for a future iteration of this problem.

We bootstrap on the analytical model developed in \cite{MortezaLAA} and further modify it to consider the different sensing duration of Wi-Fi and LTE-LAA for coexistence throughput to explore the issue of fair coexistence, first by considering the 3GPP definition. Thereafter, we investigate notions of access and proportional fairness for coexistence and discuss their pros and cons relative to the 3GPP notion. The results are illustrated for different classes of LTE-LAA traffic. The novel contributions of this paper are: 
\begin{itemize}
    \item Modifying the original analytical framework for including the impact of different sensing durations; 
    \item Characterizing when the 3GPP notion of fairness in the coexistence is achievable.
    \item Investigating when access and proportional fairness in coexistence is achievable. 
\end{itemize}

This paper is organized as follows. Section II presents a literature review of the fairness issue in Wi-Fi and coexistence system. In Section III, the Wi-Fi and LTE-LAA MAC layer protocols are described. Section IV presents the analytical modeling of coexistence network. Section V uses the 3GPP notion of fairness to achieve throughput fairness and section VI investigate the access fairness based on the 3GPP notion of fairness. In Section VII, the proportional fairness in the coexistence system is investigated. Section VIII illustrates the effect of Wi-Fi TXOP in VHT mode on the fairness. Section IX concludes the paper.

\begin{figure*}[!htb]
\setlength{\belowcaptionskip}{-0.1in}
\begin{center}
\includegraphics[width=5.0in]{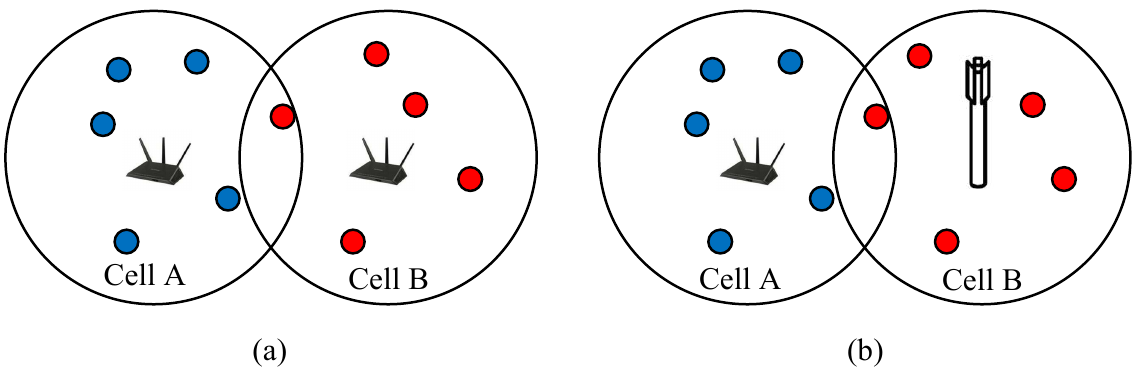}
 \caption{(a) coexistence of two Wi-Fi networks in Cell A and B. (b) coexistence of a Wi-Fi network (Cell A) with an LTE-LAA network (Cell B).}
 \label{fig: Diag}
\end{center}
\end{figure*}


\section{On Fairness in Wireless Networks}
\label{sec: literature}

\subsection{Basics of Fairness in A Wireless Network}

Fairness in a single wireless network implies sharing of a relevant resource - such as access or throughput \cite{Fairness_survey} - in an equitable manner among the nodes. The principle of fairness is invoked (often as a counterpoint to other principles such as optimality of network design) to prevent undesirable effects such as starvation or redundant allocation to a station. Fairness has been typically captured in several metrics - such as max-min and proportional fairness \cite{Fairness_survey}.  


Max-min and proportional fairness have been well-studied in the context of WLANs \cite{WLAN_Fairness2005,WLAN_Fairness2008,Fairness_11e} governed by a prototypical time-sharing MAC. Proportional fairness is well-known to achieve {\em airtime fairness} in rate-heterogeneous Wi-Fi networks. These explore the choice of Wi-Fi DCF system parameters like data rate and TXOP to achieve throughput fairness. In max-min fairness, the stations seek equal share of the available bandwidth. However, stations with different data rates or TXOP consume different airtimes which means that the max-min bandwidth fair is not airtime fair. The standard CSMA/CA in IEEE 802.11 attempts to achieve max-min fairness in bandwidth usage. As pointed out in \cite{Fairness_Heusse}, the aggregate throughput in a rate heterogeneous WLAN is thus determined by stations with the lowest data rates, leading to the `bandwidth anomaly'.

A significant observation is that proportional fairness is achieved when the fraction of airtime usage by the stations are equal \cite{WLAN_Fairness2005}. 
In such a proportional-fair WLAN, the throughput of each station is independent of their respective data rates. In \cite{WLAN_Fairness2008}, the aggregate throughput of a proportional fair WLAN 802.11b  deployment (consisting of a network of access points) was shown to as much as 2.3 times that of the max-min fair allocation.

In \cite{Fairness_11e}, the proportional fairness of IEEE 802.11e standard which uses the Enhanced Distributed Channel Access (EDCA) - whereby each node maintains a different queue for each class of traffic - is investigated. The stations transmit at different rates and have different loads for each station corresponding to their traffic classes. To achieve proportional fairness, the minimum contention window size for each access category in EDCA was optimized. When stations have different transmission rates, the optimal minimum contention window for the high data rate stations is smaller than the low data rate stations. When stations have the same traffic load, the proposed proportional fairness achieves a better throughput compared with the time-based fairness in multi-rate scenarios.

\vspace*{-0.1in}
\subsection{Fair Coexistence among Dis-similar Networks}

Defining fair coexistence of co-located dissimilar networks such as Wi-Fi (based on CSMA/CA) and unlicensed LTE (based on time division multiple access and LBT) is a challenge. 3GPP presented a definition that is {\em one-sided} - it imposes conditions on LTE-LAA to preserve WiFi throughput insensitivity, but makes no prescriptions on the total fairness of Wi-Fi and LTE-LAA networks. This is clearly different than the notion of proportional fairness which aims at fair sharing of the total (Wi-Fi and LTE-LAA) coexistence throughput. 

The FCC \cite{FCCLAA} investigation on the effect of the LTE-LAA on Wi-Fi concluded that LTE-LAA deployment would have unfair access to the channel relative to Wi-Fi and therefore decrease the latter's average throughput. Qualcomm in \cite{Qualcomm} investigated the coexistence of Wi-Fi with LTE-LAA through simulation and showed that significant throughput gain can be achieved by aggregating LTE across licensed and unlicensed spectrum; this throughput improvement does not come at the expense of degraded Wi-Fi performance and both technologies can fairly share the unlicensed spectrum.

In \cite{LTEU_fairLeith1, LTEU_fairLeith2}, the fairness of Wi-Fi and LTE-LAA, as well as Wi-Fi and LTE-U with CSAT, was investigated; it was shown that when optimally configured,  proportionally-fair deployments involving LTE-LAA and LTE-U  are capable of providing the same level of fairness to Wi-Fi. In \cite{LTEWiFi_Fair}, fairness in the coexistence of Wi-Fi/LTE-LAA based on the 3GPP criterion is investigated through an event-based system simulator. Simulation results show that the choice of LBT parameters for LTE-LAA is essential in achieving such fairness.

\begin{figure*}[!htb]
\setlength{\belowcaptionskip}{-0.1in}
\begin{center}
\includegraphics[width=5.0in]{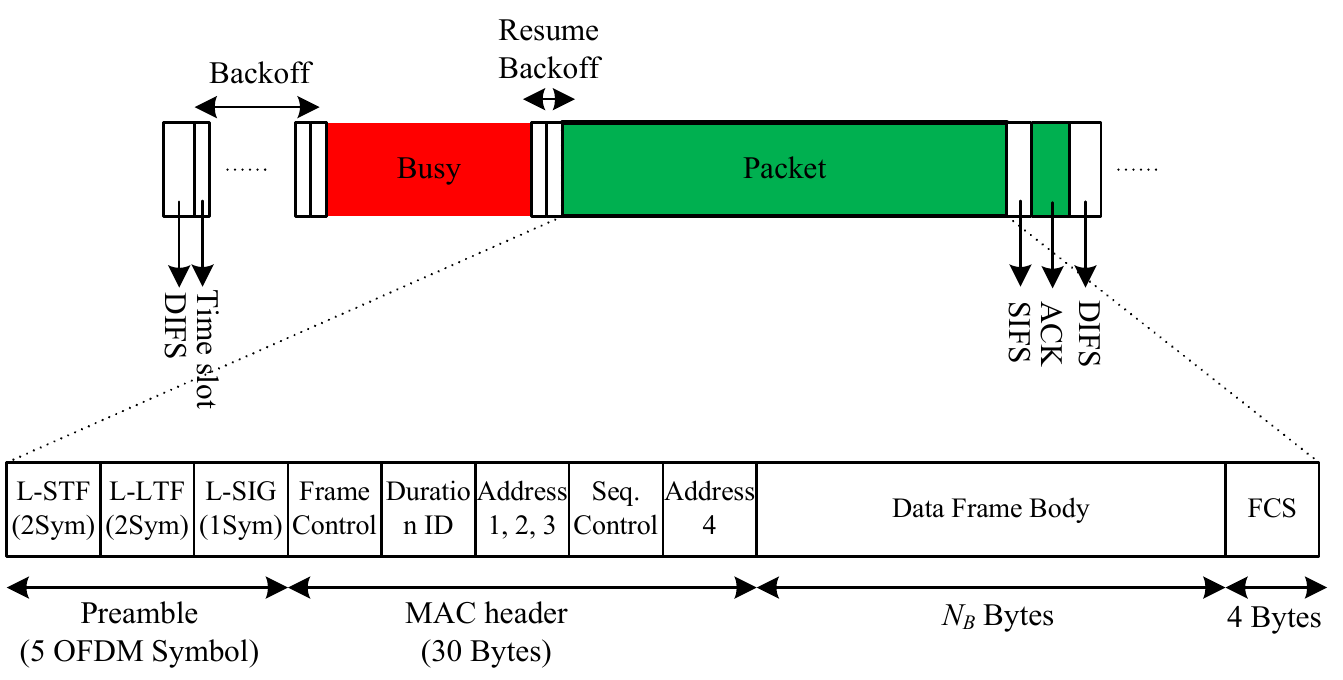}
 \caption{Wi-Fi CSMA/CA contention and packet transmission}
 \label{fig: WiFitime}
\end{center}
\end{figure*}


\section{Coexistence of LTE-LAA and Wi-Fi: MAC Protocol Mechanisms}

In this section, the MAC protocol of Wi-Fi and LTE-LAA is presented. While LTE-LAA uses LBT to mimic Wi-Fi DCF, there are some key differences which are highlighted, as these have a significant bearing on the respective channel access. 

\subsection{Wi-Fi DCF}
\label{sec: WiFiDCF}

\begin{figure}[!htb]
\setlength{\belowcaptionskip}{-0.1in}
\begin{center}
\includegraphics[width=2.7in]{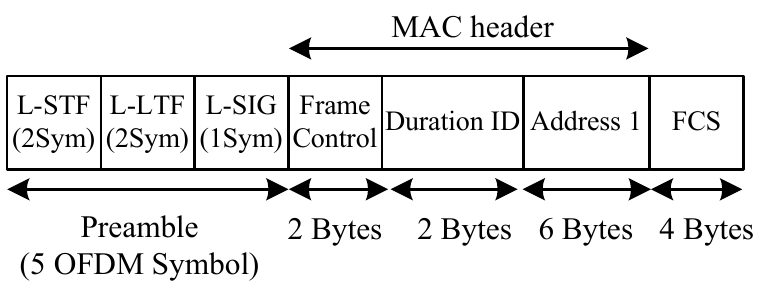}
 \caption{Wi-Fi ACK frame structure}
 \label{fig: ACKframe}
\end{center}
\end{figure}

The Wi-Fi MAC distributed coordination function (DCF) employs CSMA/CA \cite{std80211} as illustrated in Fig.~\ref{fig: WiFitime} that is explained in the following. Each node attempting transmission must first ensure that the medium has been idle for a duration of DCF Interframe Spacing (DIFS) using the ED\footnote{the ability of Wi-Fi to detect the external interference} and CS\footnote{the ability of Wi-Fi to detect and decode an incoming Wi-Fi signal preamble} mechanism. When either of ED and CS is true, the Clear Channel Assessment (CCA) is indicated as busy. If the channel is IDLE for DIFS immediately after a successful transmission, the station transmits. Otherwise, if the channel is sensed busy (either immediately or during the DIFS) or the station wants to contend after a successful transmission, the station persists with monitoring the channel until it is measured idle for a DIFS, then selects a random back-off duration (counted in units of slot time) and counts down. Specifically, a station selects a back-off counter uniformly at random in the range of $[0, 2^i W_0 - 1]$ where the value of $i$ (the back-off stage) is initialized to 0 and $W_0$ is the {\em minimum contention window} chosen initially.  Each failed transmission due to packet collision\footnote{A collision event occurs if and only if two nodes select the same back-off counter value at the end of a DIFS period.} results in incrementing the back-off stage by $1$ (binary exponential back-off or BEB) and the node counts down from the selected back-off value;  i.e. the node decrements the counter every $\sigma (\mu s)$ corresponding to a back-off slot as long as no other transmissions are detected. If during the countdown a transmission is detected, the counting is paused, and nodes continue to monitor the busy channel until it goes idle; thereafter the medium must remain idle for a further DIFS period before the back-off countdown is resumed. Once the counter hits zero, the node transmits a packet. Any node that did not complete its countdown to zero in the current round, carries over the back-off value and resumes countdown in the next round.  Once a transmission has been completed successfully, the value of $i$ is reset to 0. The maximum value of back-off stage $i$ is $m$ and it stays in $m$-th stage for one more unsuccessful transmission, i.e. the retry limit is 1. If the last transmission was unsuccessful, the node drops the packet and resets the back-off stage to $i=0$. If a unicast transmission is successful, the intended receiver will transmit an Acknowledgment frame (ACK) after Short Interframe Spacing (SIFS) duration post successful reception; the ACK frame structure is shown in Fig.~\ref{fig: ACKframe} which consists of preamble and MAC header. The ACK frame chooses the highest basic data rate (6 Mbps, 12 Mbps, or 24 Mbps) for transmitting the MAC header which is smaller than the data rate used for data transmission. 

\subsection{LTE-LAA LBT}
\label{sec: LAA}

\begin{figure*}[!htb]
\setlength{\belowcaptionskip}{-0.1in}
\begin{center}
\includegraphics[width=5.4in]{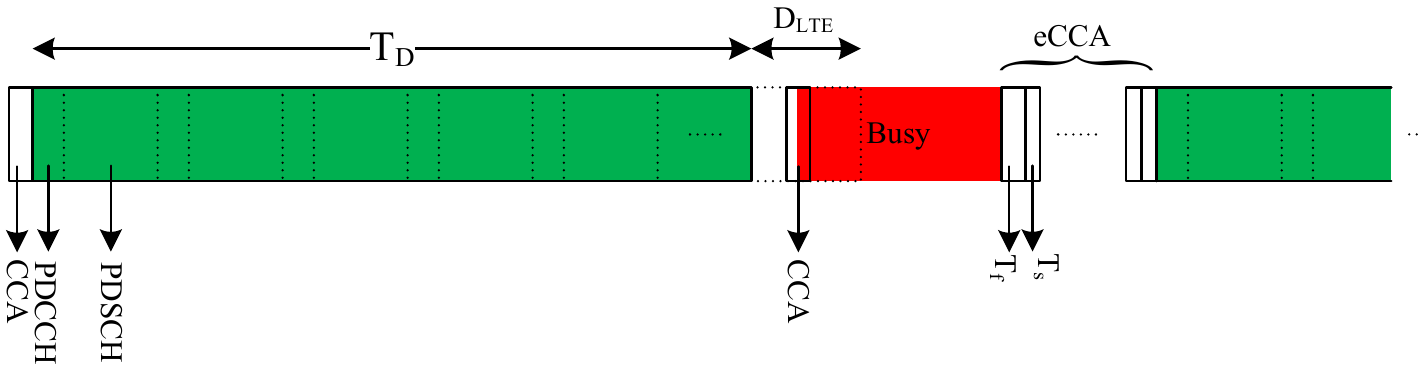}
 \caption{LTE-LAA LBT contention with CCA/eCCA and LTE subframe structure}
 \label{fig: LAAtime}
\end{center}
\end{figure*}

LTE-LAA follows LBT approach for coexistence with Wi-Fi \cite{ETSILAA17} which follows CSMA/CA with the following key differences as illustrated in Fig.~\ref{fig: LAAtime}: \\ (a) LTE-LAA performs a CCA check using ``energy detect'' (CCA-ED) where it observes the channel for the {\em defer period} ($T_d$). The $T_d$ depends on the access priority class for downlink (DL) and uplink (UL) as defined in Table~\ref{table: LAAclass1}; in class 1 and 2 the DL has the priority to UL (because the $T_d$ is smaller in the downlink for class 1 and 2). There is no CS in LTE-LAA like Wi-Fi for performing preamble detection. If channel sensed idle and the current transmission is not immediately after a successful transmission, the LTE-LAA node starts transmission; if sensed busy, it reverts to extended CCA (eCCA) whereby it senses and defers until the channel is idle for $T_d$, and then performs the exponential back-off similar to DCF (selects a back-off counter and decrements the back-off counter every slot time $T_s=9$ $\mu s$). \\ (b) As illustrated in Table~\ref{table: LAAclass1}, LTE-LAA identifies 4 channel access priority classes for both UL and DL with different minimum and maximum contention window size. \\ (c) whenever a collision happens, the back-off number is selected randomly from doubled contention window size for retransmission (i.e., $[0, 2^i W'_0 - 1]$, where $i$ is the retransmission stage for selecting the contention window size). When $i$ exceeds the maximum retransmission stage $m'$, it stays at maximum window size for $e_l$ times ($e_l$ is the retry limit after reaching to $m'$) where the $e_l$ is selected from the set of values $\{1,2,...,8\}$; then, $i$ resets to 0. \\ (d) when an LTE-LAA eNB gets access to the channel, it is allowed to transmit packets for a TXOP ($T_D$ in Fig.~\ref{fig: LAAtime}) duration of up to 10 $ms$ when known a-priori that there is no coexistence node, otherwise up to 8 $ms$ for DL and 6 $ms$ for UL.\\ (e) The minimum resolution of data transmission length in LTE-LAA is one subframe (i.e., 1 ms) and LTE-LAA transmits the subframe per $D_{LTE}=$ 0.5 $ms$ slot boundaries; \\ (f) After the maximum transmission time, if data is available at the LTE-LAA buffer, it should perform the eCCA for accessing the channel. Literally, the LTE-LAA contend and access the channel in Wi-Fi time slot resolution ($T_s$) but after accessing the channel transmits the frames in 0.5 ms resolution. 

LTE-LAA uses the base LTE subframe structure, i.e. the subframes of 1 ms duration comprising two $0.5$ ms slots. Each subframe consists of 14 OFDM symbols as in Fig.~\ref{fig: LAAtime}, of which 1 to 3 are Physical Downlink/Uplink Control Channel (PDCCH/PUCCH) symbols and the rest are Physical Downlink/Uplink Shared Channel (PDSCH/PUSCH) data. LTE-LAA start transmissions synchronized with slot boundaries, for (at least) one subframe duration. After transmission, the intended receiver (or receivers) transmits the ACK via the {\em licensed band} if decoding is successful.  

Finally, a Resource Block (RB) is the smallest unit of radio resource which can be allocated to a UE, equal to 180 kHz bandwidth over a Transmission Time Interval (TTI) of one subframe (1 ms). Each RB of 180 kHz bandwidth contains 12 sub-carriers, each with 14 OFDM symbols, equaling 168 Resource Elements (REs). Depending upon the modulation and coding schemes (QPSK, 16-QAM, 64-QAM), each symbol or resource element in the RB carries 2, 4 or 6 bits per symbol, respectively. In the LTE system with 20 MHz bandwidth, there are 100 RBs available. 

\begin{table}
\caption{LTE-LAA LBT parameters per class for downlink (DL) and uplink (UL) transmission.}
\label{table: LAAclass1}
\begin{center}
\begin{tabular}{|c|c|c|c|c|c|}
\hline 
Access Priority Class \# & $T_d$ & $W'_0$ & $m'$ & TXOP ($T_D$) \\ 
\hline
\hline
1 - DL & 25 $\mu s$ & 4 & 1 & 2 $ms$ \\ 
 \hline
2 - DL & 25 $\mu s$ & 8 & 1 & 3 $ms$ \\ 
 \hline
3 - DL & 43 $\mu s$ & 16 & 2 & 8 $ms$ or 10 $ms$\\
 \hline
4 - DL & 79 $\mu s$ & 16 & 6 & 8 $ms$ or 10 $ms$  \\
\hline
1 - UL & 34 $\mu s$ & 4 & 1 & 2 $ms$ \\ 
 \hline
2 - UL & 34 $\mu s$ & 8 & 1 & 3 $ms$ \\ 
 \hline
3 - UL & 43 $\mu s$ & 16 & 2 & 6 $ms$ or 10 $ms$\\
 \hline
4 - UL & 79 $\mu s$ & 16 & 6 & 6 $ms$ or 10 $ms$  \\
 \hline
\end{tabular}
\end{center}
\end{table}


\section{Throughput Modeling of Wi-Fi and LTE-LAA with different AIFS}
\label{sec: TputCal}

We first summarize the analytical modeling of coexistence system proposed in \cite{MortezaLAA} and extend it to include the effect of different sensing regimes prior to backoff (i.e. DIFS for Wi-Fi DCF and $T_d$ for LTE-LAA) on throughput. We assume that Wi-Fi and LTE-LAA nodes have saturation buffers (full-load), there are $n_w$ Wi-Fi stations (1 AP and $n_w-1$ UEs) and $n_l$ LTE-LAA stations (1 eNB and $n_l-1$ UEs) in the network that transmit on both UL and DL. Both Wi-Fi and LTE-LAA use a common 20 MHz channel. We consider one contention class for Wi-Fi network and 4 contention classes (corresponding to 4 priority classes for different traffic types) for LTE-LAA as suggested by 3GPP in Table \ref{table: LAAclass1}. However, in any specific scenario, all the LTE-LAA stations are assumed to belong to the {\em same} contention class to simplify the analysis and any mixed class scenario is deferred for future work. 

The probability that a Wi-Fi node access to the channel to transmit in a time slot (Wi-Fi access probability) is calculated as \cite{MortezaLAA},
\begin{equation}
\begin{split}
&\tau_w = \sum_{j=0}^{m+1}b_{j,0}\\&=\frac{2}{W_0\left(\frac{(1-(2P_{cw})^{m+1})(1-P_{cw})+2^{m}\left(P_{cw}^{m+1}-P_{cw}^{m+2}\right)(1-2P_{cw})}{(1-2P_{cw})(1-P_{cw}^{m+2})}\right) +1},
\label{eq: tauw}
\end{split}
\end{equation}
where $P_{cw}$ is the collision probability of Wi-Fi nodes. The probability that a LTE-LAA node accessing the channel to transmit in a time slot is calculated as \cite{MortezaLAA},
\begin{equation}
\begin{split}
&\tau_l = \sum_{j=0}^{m'+e_l}b_{j,0}=\\
&\frac{2}{W_0'\left(\frac{(1-P_{cl})(1-(2P_{cl})^{m'+1})}{(1-2P_{cl})(1-P_{cl}^{m'+e_l+1})}+2^{m'}\frac{P_{cl}^{m'+1}-P_{cl}^{m'+e_l+1}}{1-P_{cl}^{m'+e_l+1}}\right)+1},
\label{eq: taul}
\end{split}
\end{equation}
where $P_{cl}$ is the collision probability of LTE-LAA nodes. 

\begin{figure}[!htb]
\setlength{\belowcaptionskip}{-0.1in}
\begin{center}
\includegraphics[width=3.2in]{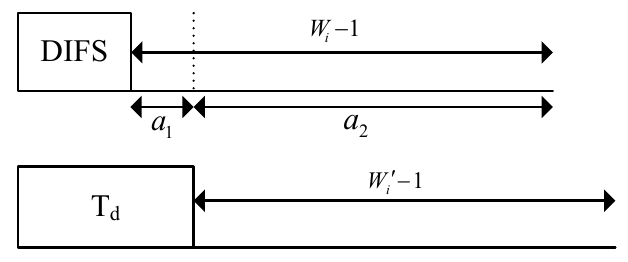}
 \caption{The contention periods when two networks with different initial sensing time accessing the channel.}
 \label{fig: Area}
\end{center}
\end{figure}

The discrepancy due to nodes with different sensing periods before backoff, e.g. Wi-Fi sensing the channel for DIFS and LTE-LAA for $T_d$ both DL and UL (we assume the UL and DL LTE-LAA have the same $T_d$ under the same priority class), leads to disparate contention as explained next and illustrated in Fig.~\ref{fig: Area}. Clearly, a smaller subset (Wi-Fi nodes only) with smaller sensing period are contending in the duration marked $a_1$ as compared to $a_2$ where both Wi-Fi and LTE-LAA nodes are in the backoff. The scenario depicted corresponds to sensing period of the Wi-Fi being smaller than LTE-LAA sensing duration, which is true for priority classes 3 and 4, per 3GPP specifications. Hence, the collision probability in $a_1$ will be smaller than that in $a_2$.

The probability of collision of  Wi-Fi with at least one of the other remaining stations ($n_w-1$ Wi-Fi) given that the transmission occurs in the first contention duration $a_1$ is calculated as:
\begin{equation}
P_{cw,1} =1-(1-\tau_w)^{n_w-1},
\label{eq: Pw1}
\end{equation}
In the second duration ($a_2$) both Wi-Fi and LTE-LAA stations are contending, so the probability of Wi-Fi collision with other remaining stations ($n_w-1$ Wi-Fi or $n_l$ LTE-LAA) given the transmission in the second contention period is calculated as:
\begin{equation}
P_{cw,2} =1-(1-\tau_w)^{n_w-1} (1-\tau_l)^{n_l}.
\label{eq: Pw2}
\end{equation}

The collision probability for an LTE-LAA stations with any other remaining stations contending in the second duration $a_2$ (there is no contention for LTE-LAA in the first period) is calculated as,
\begin{equation}
P_{cl} =1-(1-\tau_l)^{n_l-1} (1-\tau_w)^{n_w},
\label{eq: Pl}
\end{equation}
where similarly $P_{cl}$ is coupled to both Wi-Fi and LTE-LAA via $\tau_w$ and $\tau_l$. 

\begin{figure}[!htb]
\setlength{\belowcaptionskip}{-0.1in}
\begin{center}
\includegraphics[width=3.4in]{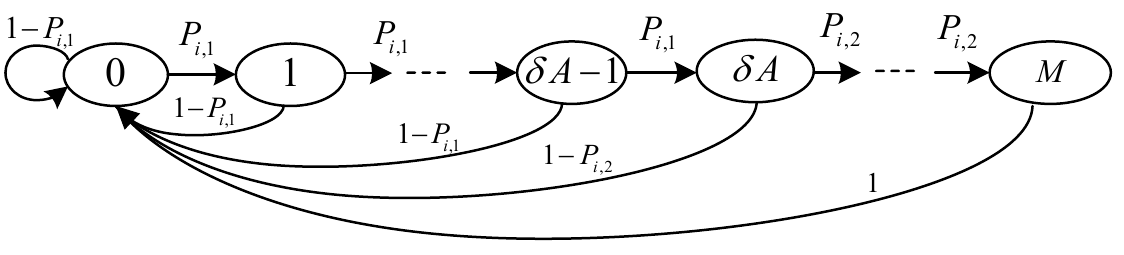}
 \caption{Markov model to represent the transition in the areas}
 \label{fig: MMarea}
\end{center}
\end{figure}

To further progress on the probability of collision computation in the first and second duration as above, we need a Markov model shown in Fig.~\ref{fig: MMarea} that represents the contention in the two durations. Each state corresponds to a time slot, and $\delta A=\frac{T_d-DIFS}{\sigma}$ is the number of time slot equivalents for the duration $a_1$ when $T_d>DIFS$, and $M=\min(W_m-1,W'_{m'}-1+\delta A)$ is the total possible number of slots for backoff contention which is limited by the maximum backoff window value. The states from 0 to $\delta A-1$ are the slots in the first contention period. $P_{i,1}$,  the idle probability in the first contention period is thus 
\begin{equation}
P_{i,1} =(1-\tau_w)^{n_w},
\label{eq: Pi1}
\end{equation}
and similarly the idle probability in the second contention period is given by 
\begin{equation}
P_{i,2} =(1-\tau_w)^{n_w}(1-\tau_l)^{n_l}.
\label{eq: Pi2}
\end{equation}

The normalization condition is 
\begin{equation}
\sum_{k=0}^{k=M}c_k=1,
\label{eq: norm}
\end{equation}
where $c_k$ for the first contention period is 
\begin{equation}
c_k=c_0 (P_{i,1})^k,
\label{eq: ck1}
\end{equation}
and for the second contention period, 
\begin{equation}
c_k=c_0 (P_{i,1})^{\delta A} (P_{i,2})^{k-\delta A}.
\label{eq: ck2}
\end{equation}

Using (\ref{eq: norm}), (\ref{eq: ck1}), and (\ref{eq: ck2}) $c_0$ is calculated as:
\begin{equation}
c_0=\left[\frac{1-(P_{i,1})^{\delta A+1}}{1-P_{i,1}}+(P_{i,1})^{\delta A}P_{i,2} \frac{1-(P_{i,2})^{M-\delta A}}{1-P_{i,2}} \right]^{-1}.
\label{eq: c0}
\end{equation}

Thus, the probability of contention in a random slot in the first period can be calculated as:
\begin{equation}
P_{a,1}=\sum_{k=0}^{\delta A-1}c_k=c_0\frac{1-(P_{i,1})^{\delta A}}{1-P_{i,1}},
\label{eq: pa1}
\end{equation}
and the probability of contention in a random slot in the second period is: 
\begin{equation}
P_{a,2}=1-P_{a,1}.
\label{eq: pa2}
\end{equation}

The total collision probability of Wi-Fi based on the two contention period is the weighted average: 
\begin{equation}
P_{cw} = P_{a,1}P_{cw,1}+P_{a,2}P_{cw,2}.
\label{eq: Pw}
\end{equation}

Therefore, $P_{cw}$ is coupled to both Wi-Fi and LTE-LAA nodes via $\tau_w$ and $\tau_l$.  
To compute the $P_{cw}$, $P_{cl}$, $\tau_w$, and $\tau_l$ for the coexistence scenario, we must jointly solve (\ref{eq: tauw}), (\ref{eq: taul}), (\ref{eq: Pl}), (\ref{eq: pa1}), and (\ref{eq: Pw}).

The transmission probability of Wi-Fi (that at least one of the $n_w$ stations transmit a packet during a time slot) is:
\begin{equation}
P_{trw} =1-(1-\tau_w)^{n_w},
\label{eq: Ptrw}
\end{equation}
and similarly the transmission probability of  LTE-LAA is:
\begin{equation}
P_{trl} =1-(1-\tau_l)^{n_l}.
\label{eq: Ptrl}
\end{equation}

The successful transmission of a Wi-Fi node is the event that exactly one of the $n_w$ stations makes a transmission attempt given that at least one of the Wi-Fi stations transmit:
\begin{equation}
P_{sw} =\frac{n_w \tau_w (1-\tau_w)^{n_w-1}}{P_{trw}}.
\label{eq: Psw}
\end{equation}
Similarly the successful transmission probability of LTE-LAA is:
\begin{equation}
P_{sl} =\frac{n_l \tau_l (1-\tau_l)^{n_l-1}}{P_{trl}}.
\label{eq: Psl}
\end{equation}

To compute the average throughput of Wi-Fi, we need the average time duration for a successful transmission and collision event, respectively, given by: 
\begin{equation}
\begin{split}
T_{sw} = &\text{PhyH}+\text{MACH}+\text{Psize}+\text{SIFS}+\text{ACK}+\text{DIFS} \\
T_{cw} = &\text{PhyH}+\text{MACH}+\text{Psize}+\text{DIFS} \\
\end{split}.
\label{eq: TsTc}
\end{equation}

The average time duration of successful transmission event and collision event for LTE-LAA are:
\begin{equation}
\begin{split}
T_{sl} &= T_D+D_{LTE}\\
T_{cl} &= T_D+D_{LTE},
\end{split}
\label{eq: TimeLAA}
\end{equation}
where the $T_D$ is the TXOP duration of LTE-LAA and $D_{LTE}$ is the delay for the next transmission which is one LTE slot (0.5 $ms$). After transmission for LTE-LAA TXOP duration, the transmitter waits for the ACK and then resumes channel contention for the next transmission opportunity. If an LTE eNB wins channel contention {\em before} start of next LTE slot, the LTE-LAA transmits a reservation signal to keep the channel until the end of the current LTE slot to start transmission.

The throughput of Wi-Fi is calculated as:
\begin{equation}
Tput_{w} =\frac{ \left [P_{a,1}P_{trw} P_{sw}+P_{a,2}P_{trw} P_{sw} (1-P_{trl}) \right] Psize}{ T_E }r_w,
\label{eq: tputw}
\end{equation}
where $r_w$ is the Wi-Fi data rate. $T_{E,1}$ is the average time for a transmission in the first contention period given by 
\begin{equation}
\begin{split}
T_{E,1} =(1-P_{trw})\sigma+P_{trw}P_{sw}T_{sw}+P_{trw}(1-P_{sw})T_{cw},
\end{split}
\label{eq: TE1}
\end{equation}
Similarly, $T_{E,2}$ is the average time for a transmission in the second contention period: 
\begin{equation}
\begin{split}
T_{E,2} &=(1-P_{trw})(1-P_{trl}) \sigma+P_{trw}P_{sw}(1-P_{trl})T_{sw}\\&+P_{trl}P_{sl}(1-P_{trw})T_{sl}+P_{trw}(1-P_{sw})(1-P_{trl})T_{cw}\\&+P_{trl}(1-P_{sl})(1-P_{trw})T_{cl}+
(P_{trw}P_{sw}P_{trl}P_{sl} \\&+ P_{trw}P_{sw}P_{trl}(1-P_{sl})+ P_{trw}(1-P_{sw})P_{trl}P_{sl}  \\&+ P_{trw}(1-P_{sw})P_{trl}(1-P_{sl}))T_{cc},
\end{split}
\label{eq: TE2}
\end{equation}
where $T_{cc}$ is the average time interval for collision between Wi-Fi and LTE-LAA, determined by the larger value between $T_{cw}$ and $T_{cl}$.
The $T_{E}$ which is the total expected average time of a transmission during first or second contention period is then 
\begin{equation}
\begin{split}
T_{E} =P_{a,1}T_{E,1}+P_{a,2}T_{E,2}.
\end{split}
\label{eq: TEt}
\end{equation}

Similarly the throughput of the LTE-LAA is calculated as,
\begin{equation}
Tput_{l} =\frac{P_{a,2}P_{trl} P_{sl} (1-P_{trw}) \frac{13}{14} T_D}{ T_E }r_l,
\label{eq: tputl}
\end{equation}
where $\frac{13}{14} T_D$ is the fraction of the LTE-LAA TXOP in which the data is transmitted, i.e. 1 PDCCH symbol in a subframe with 14 OFDM symbols, and $r_l$ is the LTE-LAA data rate.


\section{3GPP Fairness}
\label{3GPPfair}

In this section, we explore the 3GPP notion of fairness as defined in \cite{3GPP_TR} ``the LAA design should target fair coexistence with existing Wi-Fi networks to not impact Wi-Fi services more than an additional Wi-Fi network on the same carrier with respect to throughput and latency'' via analytical modeling. We assume that the $T_d$ value for class 1 and 2 of UL and DL in Table~\ref{table: LAAclass1} is same as Wi-Fi DIFS duration 34 $\mu s$. We emphasize again that the 3GPP `fairness' definition is {\em one-sided}, i.e. only imposes a condition on Wi-Fi network operation, without regard to LTE-LAA. 

\subsection{Per User Throughput Fairness}

In light of the above, we propose `per user throughput' as the appropriate metric for investigating 3GPP fairness. Let us consider two network scenarios: in the first, 2 overlapping co-channel Wi-Fi cells, i.e. 2 Wi-Fi AP and $N-2$ Wi-Fi UE's that are coexisting (which from a model perspective is equivalent to a single Wi-Fi network of $N$ stations) and in the second, $n_w$ Wi-Fi stations (one Wi-Fi AP and $n_w-1$ Wi-Fi UE's) coexisting with a co-channel LTE-LAA network with $n_l$ stations (one LTE-LAA eNB and $n_l-1$ LTE-LAA UE's) in which $n_w=n_l=N/2$.
To achieve per user throughput fairness, the throughput of a station in Wi-Fi only network with $N$ stations should be equal to the throughput of Wi-Fi stations in the coexistence network, i.e., 
\begin{equation}
\frac{Tput_{wo}}{N}=\frac{Tput_{w}}{n_w},
\label{eq: TF}
\end{equation}
where $Tput_{w}$ is the Wi-Fi throughput in coexistence and $Tput_{wo}$ is the Wi-Fi throughput in Wi-Fi only network which is calculated as:
\begin{equation}
\begin{split}
&Tput_{wo} =\frac{P_{tr} P_{s}Psize}{(1-P_{tr})\sigma+P_{tr}(1-P_{s})T_{cw}+P_{tr}P_{s}T_{sw}}r_w, \\
&P_{tr} =1-(1-\tau)^{N},\\
&P_{s} =\frac{N \tau (1-\tau)^{N-1}}{P_{tr}}.
\label{eq: tputwo}
\end{split}
\end{equation}
where (\ref{eq: tputwo}) depends on $W_0$, $m$, $N$, $Psize$, and $r_w$ as the parameters of Wi-Fi network. The $Tput_{w}$ in (\ref{eq: tputw}) depends on $W'_0$, $m'$, $W_0$, $m$, $\delta A$, $n_w$, $n_l$, $Psize$, $T_D$ (LTE-LAA TXOP), and $r_w$. We choose one of the parameters of the LTE-LAA network $T_D$ to optimize for achieving fairness via
\begin{equation}
\begin{split}
&\underset{T_D}{\min}~ \left | \frac{Tput_{wo}}{N}-\frac{Tput_{w}}{n_w} \right|,\\
& s.t.~~0 < T_D < 6~\text{ms}.
\end{split}
\label{eq: Opt2}
\end{equation}
that ensures that the 3GPP definition of throughput fairness in coexistence is met. 

\subsection{Numerical Results of 3GPP Fairness}
\label{sec: Num3GPPFair}

An analytical model for Wi-Fi throughput estimation in a coexistence system was already developed and validated in our prior work \cite{MortezaLAA}. In this work, we extended the prior model to consider different sensing period and use it to investigate 3GPP (and other) notions of fairness. We consider the 4 LTE-LAA priority classes as presented in Section \ref{sec: LAA} in coexistence with Wi-Fi DCF (Wi-Fi parameters are listed in Table~\ref{table: WiFipar}) and choose the LTE-LAA parameters one at a time for all stations of the network to explore the feasibility of the fairness. 

\begin{table}
\caption{Wi-Fi and LTE-LAA parameters.}
\label{table: WiFipar}
\begin{center}
\begin{tabular}{|c|c|}
\hline 
Parameter & value \\
\hline
\hline
 $r_0$ & 6/12/24 Mbps \\
\hline
 PhyH & 20 $\mu s$ \\
 \hline
 MACH & (34 bytes)$/r_w$ $\mu s$\\
 \hline
 ACK & 20+(14 bytes)$/r_0$ $\mu s$ \\
 \hline
 $\delta$ & 0.1 $\mu s$ \\
 \hline
 $\sigma$ & 9 $\mu s$ \\
 \hline
 DIFS & 34 $\mu s$\\
  \hline
 SIFS & 16 $\mu s$\\
   \hline
 $m$ & 6\\
   \hline
 $W_0$ & 16\\
  \hline
 $N_B$ & 2048 (bytes) \\
  \hline
 Psize & $N_B/r_w$ $\mu s$\\
 \hline
 $D_{LTE}$ & 0.5 $ms$\\
 \hline
\end{tabular}
\end{center}
\end{table}

\begin{figure}[t]
\setlength{\belowcaptionskip}{-0.1in}
\centerline{\includegraphics[width=3.3in]{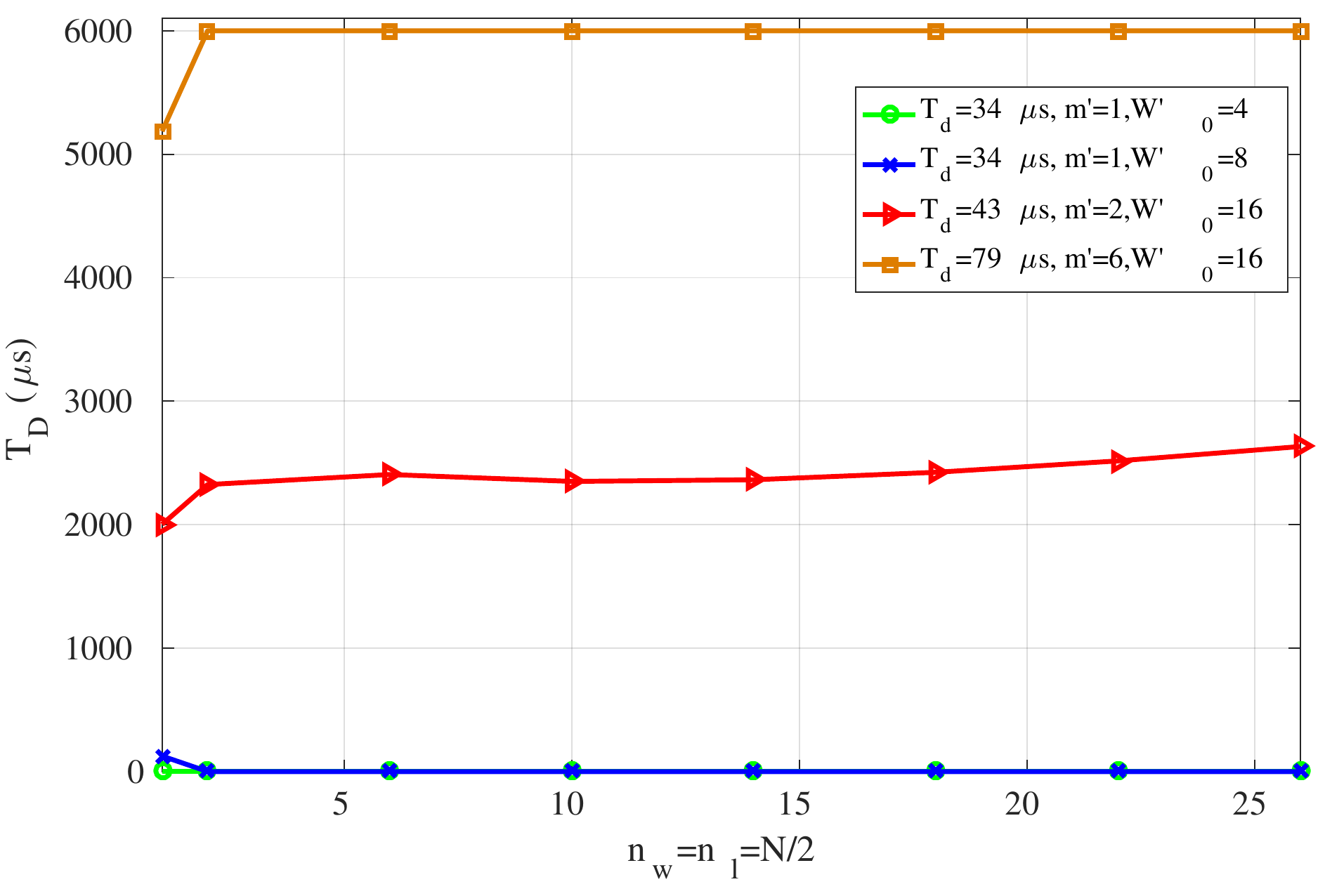}}
 \caption{The optimized LTE-LAA TXOP to achieve the throughput fairness for the 9 Mbps Wi-Fi data rate and 7.8 Mbps LTE-LAA data rate. The Wi-Fi DCF parameters are fixed as: DIFS $=34\mu s$, $m=6$, and $W_0=16$.}
 \label{fig: sim1}
\end{figure}

Fig.~\ref{fig: sim1} illustrates the optimized LTE-LAA TXOP value obtained by solving eq (\ref{eq: Opt2}) that achieves 3GPP throughput fairness. The data rate of Wi-Fi is 9 Mbps (BPSK, code rate $= 0.5$) and LTE-LAA is 7.8 Mbps (QPSK, code rate $= 0.25$) with 100 RBs of LTE-LAA corresponding to 20 MHz bandwidth, so as to equal Wi-Fi channelization. The other parameters are listed in Table~\ref{table: WiFipar}. For the cases with $m'$, $W'_0$ smaller than Wi-Fi and $T_d$ the same as Wi-Fi (priority class 1 and 2), the optimized LTE-LAA TXOP is zero (except the $n_l=1$ for priority class 2). This is because LTE-LAA gets more access to the channel and 0.5 ms of the LTE-LAA airtime is wasted for sensing, contention, and sending a reservation signal (for keeping the channel), therefore, the optimal TXOP of LTE-LAA for achieving the Wi-Fi 3GPP fairness is zero (a value smaller than 0.5 ms airtime results in Wi-Fi fairness which leaves no airtime based on eq. (\ref{eq: TimeLAA}) for LTE-LAA TXOP for data transmission). For the case $W'_0$ equal to Wi-Fi value and $T_d$ larger than Wi-Fi DIFS (priority class 3 and 4), the LTE-LAA gets lower access than Wi-Fi, the optimal LTE-LAA TXOP is expected to be larger in order to achieve fairness. We limit the LTE-LAA TXOP to 6 ms, otherwise, the optimized LTE-LAA TXOP for the priority class 4 could be larger. 

\begin{figure}[t]
\setlength{\belowcaptionskip}{-0.1in}
\centerline{\includegraphics[width=3.3in]{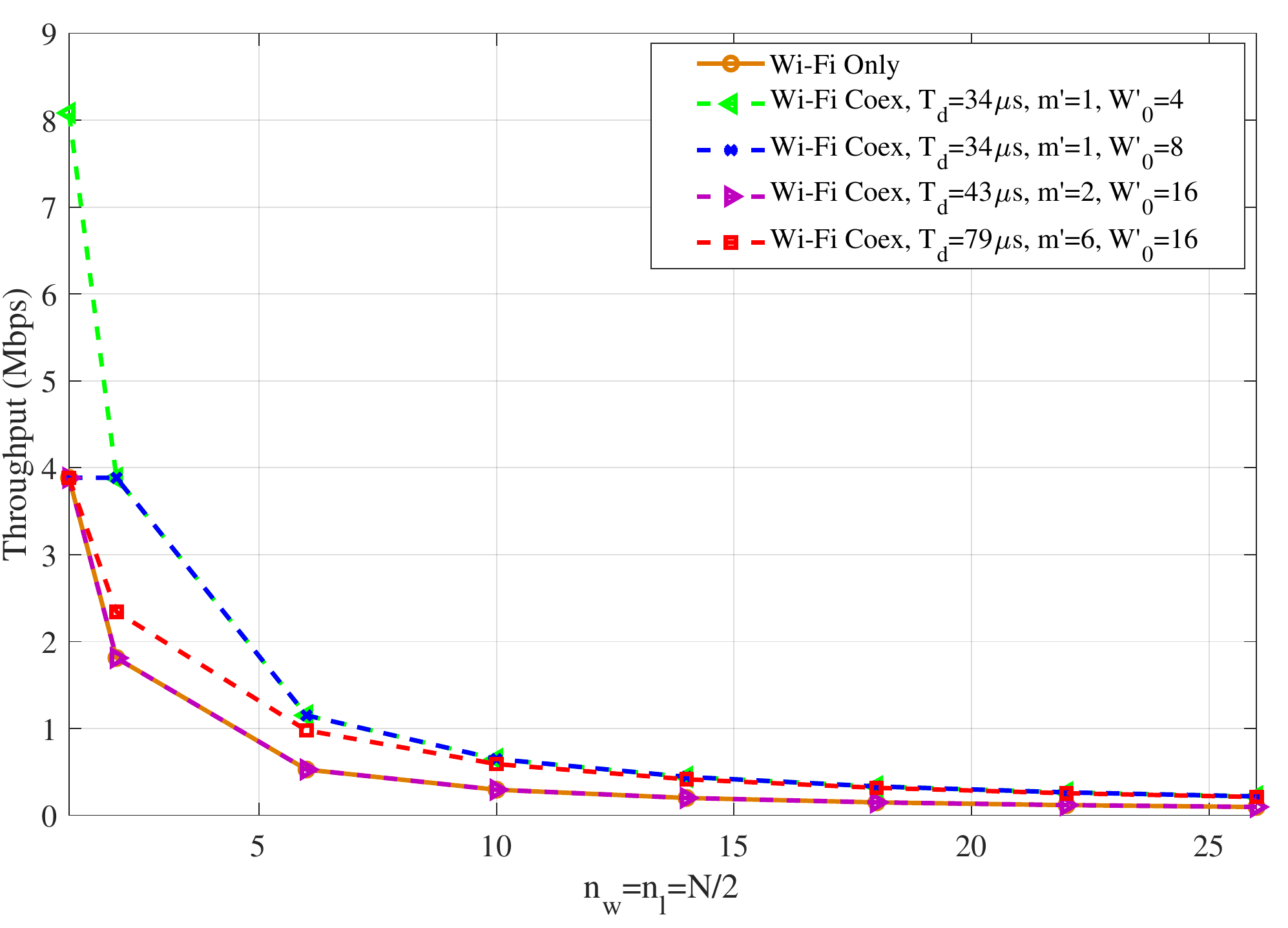}}
 \caption{The per-use throughput of Wi-Fi for the 9 Mbps Wi-Fi data rate and 7.8 Mbps LTE-LAA data rate when the Wi-Fi throughput fairness is achieved. The Wi-Fi DCF parameters are fixed as: DIFS $=34\mu s$, $m=6$, and $W_0=16$.}
 \label{fig: sim2}
\end{figure}

\begin{figure}[t]
\setlength{\belowcaptionskip}{-0.1in}
\centerline{\includegraphics[width=3.3in]{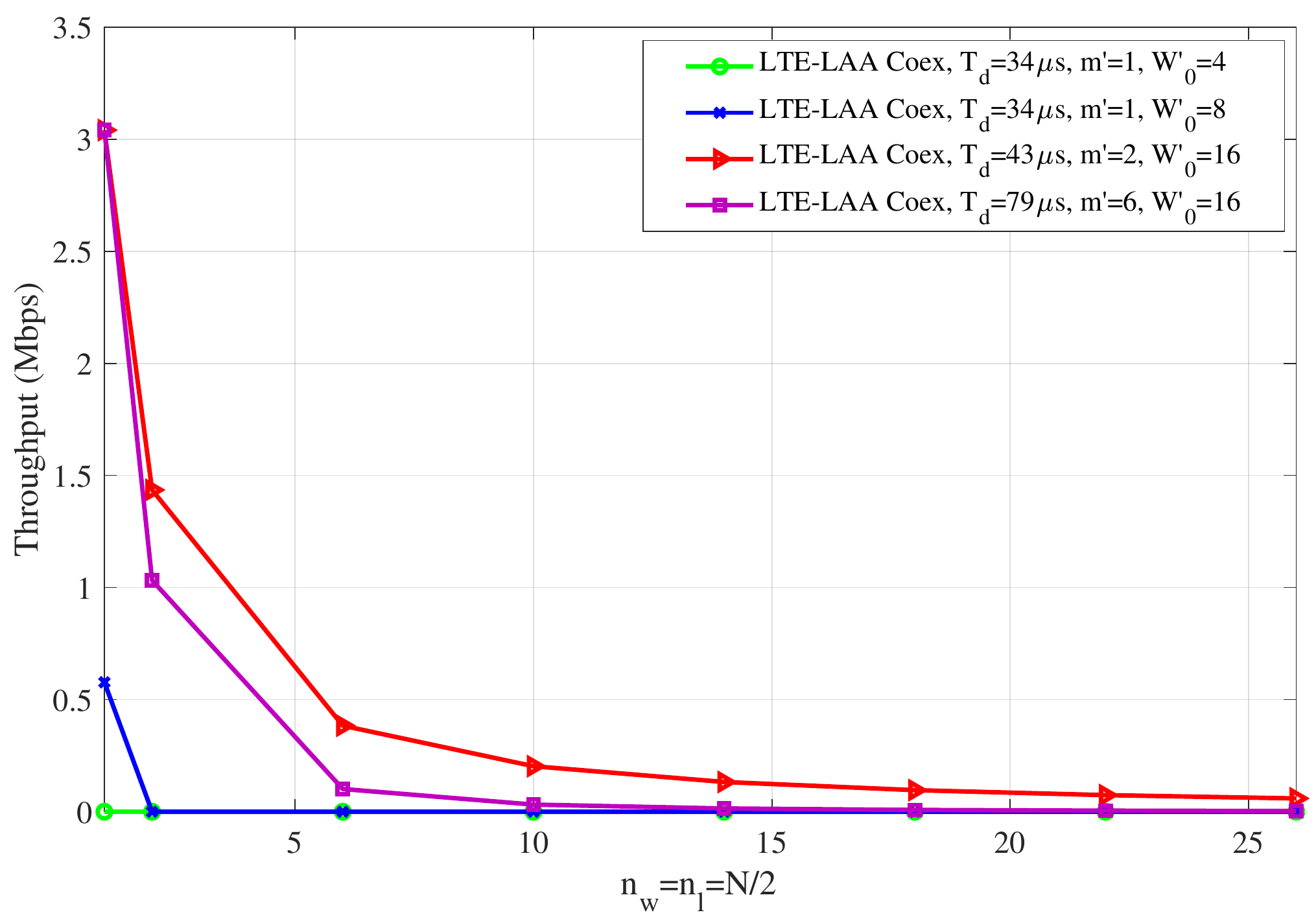}}
 \caption{The per-use throughput of LTE-LAA for the 9 Mbps Wi-Fi data rate and 7.8 Mbps LTE-LAA data rate when the 3GPP throughput fairness is achieved. The Wi-Fi DCF parameters are fixed as: DIFS $=34\mu s$, $m=6$, and $W_0=16$.}
 \label{fig: sim3}
\end{figure}

Fig.~\ref{fig: sim2} shows the per-user throughput of Wi-Fi in Wi-Fi only network (the number of Wi-Fi nodes in Wi-Fi only network is $N$ which is not shown in the Wi-Fi Only curve in this figure) and coexistence network when the 3GPP throughput fairness is achieved. For the cases that the optimized LTE-LAA TXOP is zero, contention between Wi-Fi and LTE-LAA is eliminated. In coexisting with any of the LTE-LAA priority classes, Wi-Fi achieves a higher per-user throughput than the Wi-Fi only network, due to the fact that since the target is Wi-Fi throughput fairness, LTE-LAA reduces its own throughput (or equivalently airtime) to achieve that goal. Fig.~\ref{fig: sim3} illustrates the LTE-LAA throughput in coexistence when fairness is achieved. For priority class 1 and 2, the throughput is almost zero, because of the aforementioned reasons. However, the higher priority classes achieve a higher (close to Wi-Fi) throughput. The priority class 4 has smaller throughput than class 3 because the optimized LTE-LAA TXOP is limited to 6 ms.

\begin{figure}[t]
\setlength{\belowcaptionskip}{-0.1in}
\centerline{\includegraphics[width=3.3in]{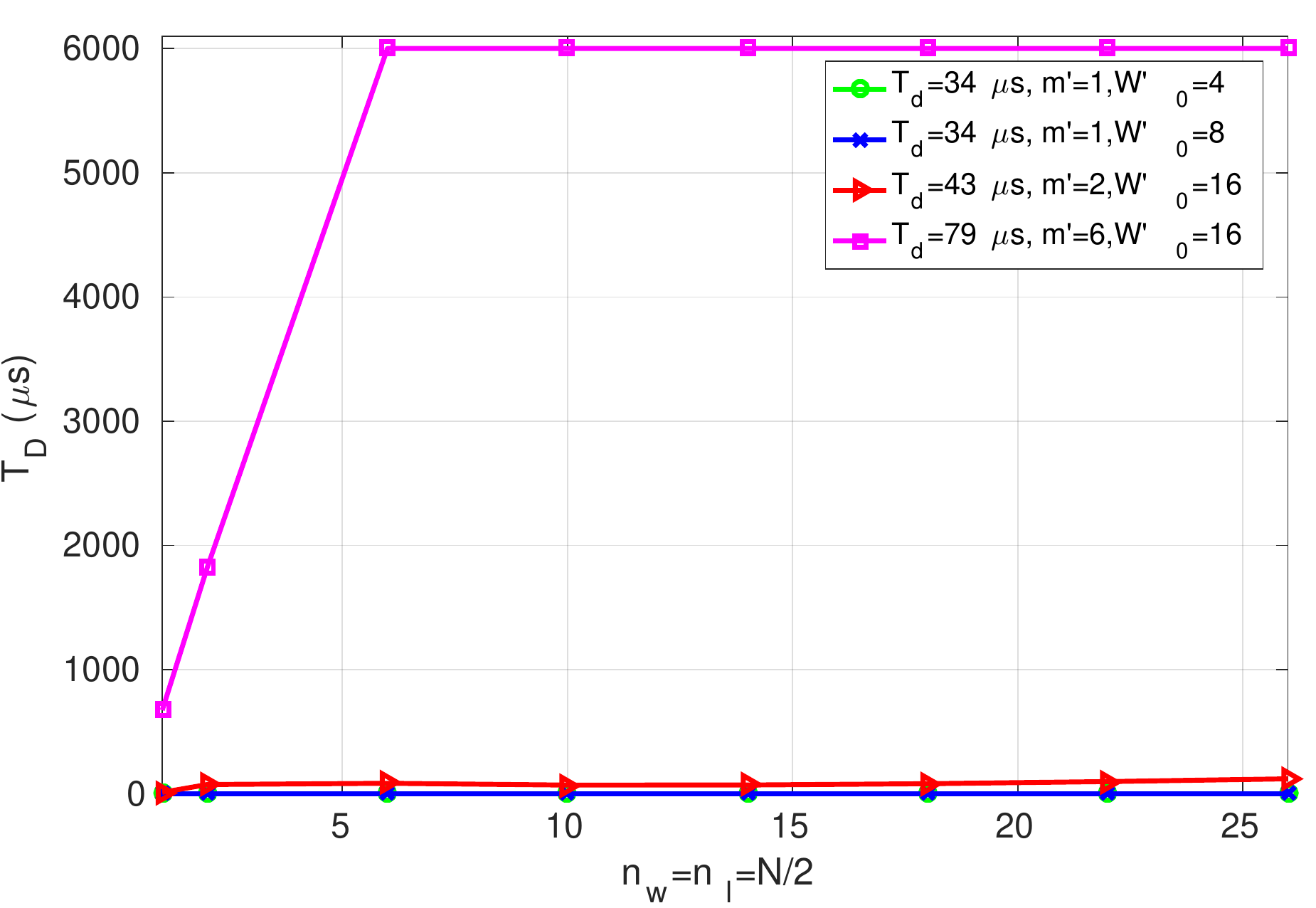}}
 \caption{The optimized LTE-LAA TXOP to achieve the throughput fairness for the 54 Mbps Wi-Fi data rate and 70.2 Mbps LTE-LAA data rate. The Wi-Fi DCF parameters are fixed as: DIFS $=34\mu s$, $m=6$, and $W_0=16$.}
 \label{fig: sim4}
\end{figure}

\begin{figure}[t]
\setlength{\belowcaptionskip}{-0.1in}
\centerline{\includegraphics[width=3.3in]{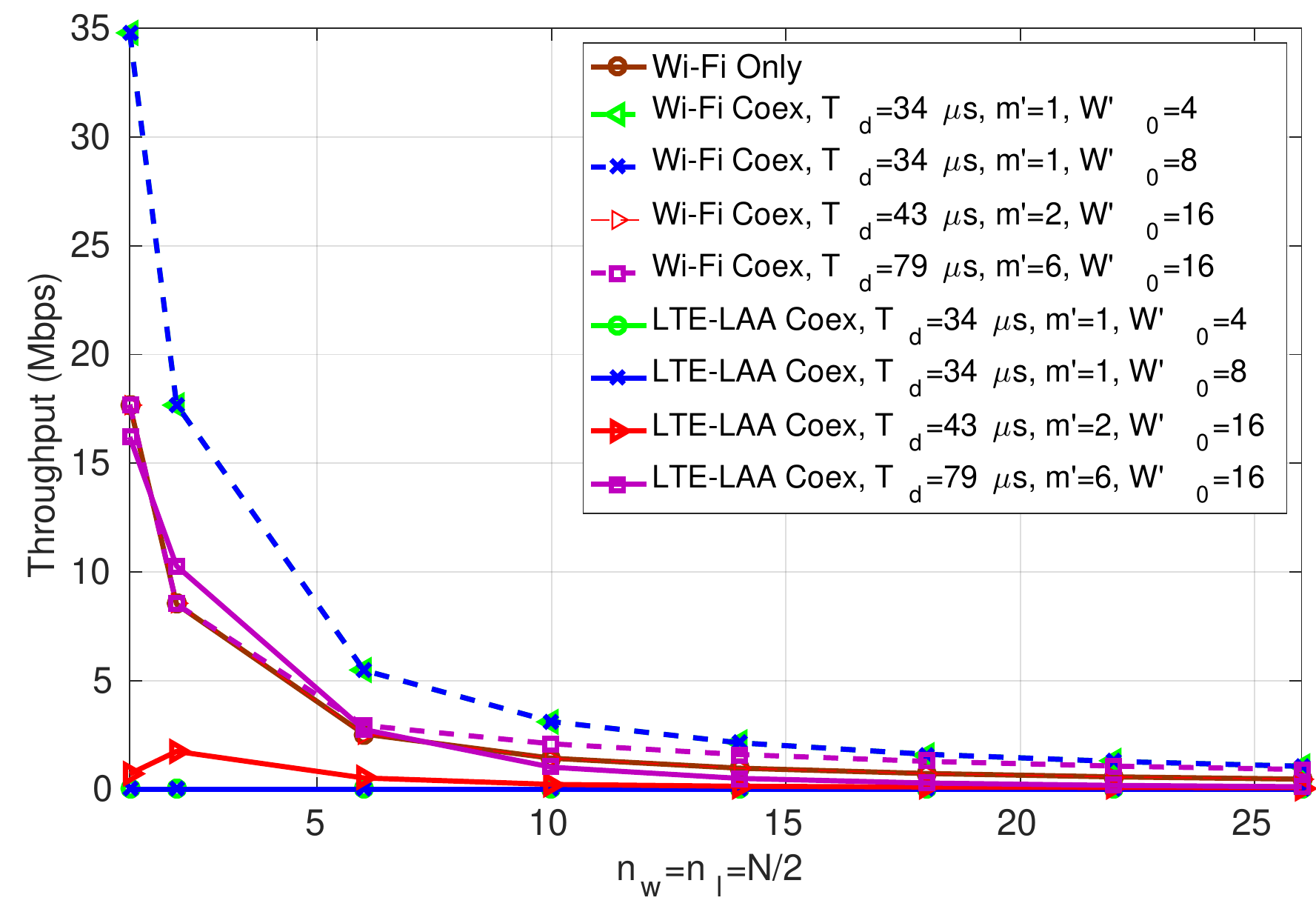}}
 \caption{The per-use throughput of Wi-Fi and LTE-LAA when the Wi-Fi throughput fairness is achieved for the 54 Mbps Wi-Fi data rate and 70.2 Mbps LTE-LAA data rate. The Wi-Fi DCF parameters are fixed as: DIFS $=34\mu s$, $m=6$, and $W_0=16$.}
 \label{fig: sim5}
\end{figure}

Fig.~\ref{fig: sim4} and Fig.~\ref{fig: sim5} illustrates the optimized LTE-LAA TXOP and achieved throughput of Wi-Fi and LTE-LAA for Wi-Fi data rate of 54 Mbps and LTE-LAA data rate of 70.2 Mbps, respectively. A similar observation with lower data rate holds with the difference that for the LTE-LAA priority class 3, the LTE-LAA TXOP is much smaller. 

\textbf{Summary:}
\begin{itemize}
    \item For the coexistence of Wi-Fi and LTE-LAA with priority class 1 and 2, Wi-Fi throughput 3GPP fairness is achieved only when TXOP and throughput of LTE-LAA are zero. Clearly, 3GPP `fairness' is not an effective definition in such scenarios.
    
    \item LTE-LAA with priority class 3 and 4 in coexistence with Wi-Fi achieves the Wi-Fi throughput fairness. However, LTE-LAA priority class 4 with lower data rate ($r_l=7.8$ Mbps) achieves a much smaller throughput than Wi-Fi in coexistence and LTE-LAA priority class 3 with higher data rate ($r_l=70.2$ Mbps) achieves a much smaller throughput than Wi-Fi in coexistence. This shows an imbalance of Wi-Fi and LTE-LAA throughput at `fair' coexistence.   
    
    \item In all of the scenarios (i.e. different priority classes) which are considered for throughput fairness in Fig.~\ref{fig: sim1}, the LTE-LAA TXOP to achieve fairness is different than the TXOP defined in \cite{ETSILAA17} as shown in the Table~\ref{table: LAAclass1}. In conclusion, \textit{only priority class 4 can use the parameters defined in \cite{ETSILAA17} and coexist fairly with Wi-Fi}.  
\end{itemize}

\section{Access Fairness}

Given the issues with the 3GPP notion of `fairness' noted above, we now seek to explore alternative {\em two-sided} fairness definitions - such as those enshrined in prior wireless networking literature such as access fairness. We first note that if the Wi-Fi frame airtime and LTE-LAA TXOP are equal and have equal data rates, access fairness is equivalent to throughput fairness. We explore when such fairness is achievable, considering exactly the same scenarios as the previous section: in the first, a network with $N$ Wi-Fi stations (2 Wi-Fi AP and $N-2$ Wi-Fi UE's in Wi-Fi only network) are coexisting and in the second, $n_w$ Wi-Fi stations (one Wi-Fi AP and $n_w-1$ Wi-Fi UE's) and $n_l$ LTE-LAA stations (one LTE-LAA eNB and $n_l-1$ LTE-LAA UE's) are coexisting in which $n_w=n_l=N/2$.
To achieve access fairness, the probability for a station to transmit at a randomly selected time slot in Wi-Fi only network with $N$ stations should equal the access probability for Wi-Fi stations in the coexistence network. This condition implies:

\begin{equation}
\tau(P)=\tau_w(P_{cw},\tau_l,P_{cl}),
\label{eq: AF}
\end{equation}
where $\tau(P)$ is defined below, the $\tau_w$ and $P_{cw}$ are defined in (\ref{eq: tauw}) and (\ref{eq: Pw}) where the $P_{cw}$ is a function of $\tau_l$ and consequently $P_{cl}$. $\tau$ - the probability that a Wi-Fi station in a Wi-Fi only network transmits in a randomly selected time slot -  is calculated considering that all Wi-Fi stations contend in one priority class with DIFS period sensing:
\begin{equation}
\begin{split}
&\tau (P) = \frac{2}{W_0\left(\frac{(1-(2P)^{m+1})(1-P)+2^{m}\left(P^{m+1}-P^{m+2}\right)(1-2P)}{(1-2P)(1-P^{m+2})}\right) +1},\\
&P =1-(1-\tau)^{N-1},
\label{eq: Ptau}
\end{split}
\end{equation}
where the $W_0$ and $m$ parameters for the Wi-Fi nodes are the same for both Wi-Fi only and coexistence network scenarios. To satisfy the access fairness in (\ref{eq: AF}), we have to solve (\ref{eq: tauw}), (\ref{eq: Pw}), (\ref{eq: taul}), (\ref{eq: Pl}), along with the (\ref{eq: Ptau}) by optimizing the LTE-LAA parameters such as the minimum contention window size and/or maximum retransmission stage in LTE-LAA. This is based on the assumption that the Wi-Fi parameters are fixed and by adapting the LTE-LAA parameters, we achieve Wi-Fi fairness in the coexistence network. 
Choosing to optimize the LTE-LAA maximum retransmission stage ($m'$) to achieve access fairness, leads to the following based on the given $W_0$, $m$, $W'_0$, $\delta A$, $N$, $n_l$, and $n_w$ parameters of Wi-Fi and LTE-LAA:
\begin{equation}
\begin{split}
&\underset{m'}{\min}~ \left | \tau(W_0,m,N)-\tau_w(W_0,m,W'_0,\delta A,n_l,n_w)
\right|,\\
& s.t.~~ m'\ge 0.
\end{split}
\label{eq: Opt}
\end{equation}

Clearly, the solution to the above is a function of the number of stations in each network ($N$, $n_w$, and $n_l$); this is investigated in the next section through numerical calculation.

As can be seen, eq. (\ref{eq: Opt}) is independent of TXOP or data rate of Wi-Fi and LTE-LAA. As already discussed, if the Wi-Fi and LTE-LAA airtime and data rate are equal, the access fairness is equivalent to the throughput fairness; so, by achieving access fairness, we are satisfying the 3GPP definition of throughput fairness as well. {\em However, in general, the two systems have different operators who do not collaborate} and hence differences in data rate or TXOP settings are to be expected.

\subsection{Numerical Results}

\begin{figure}[t]
\setlength{\belowcaptionskip}{-0.1in}
\centerline{\includegraphics[width=3.3in]{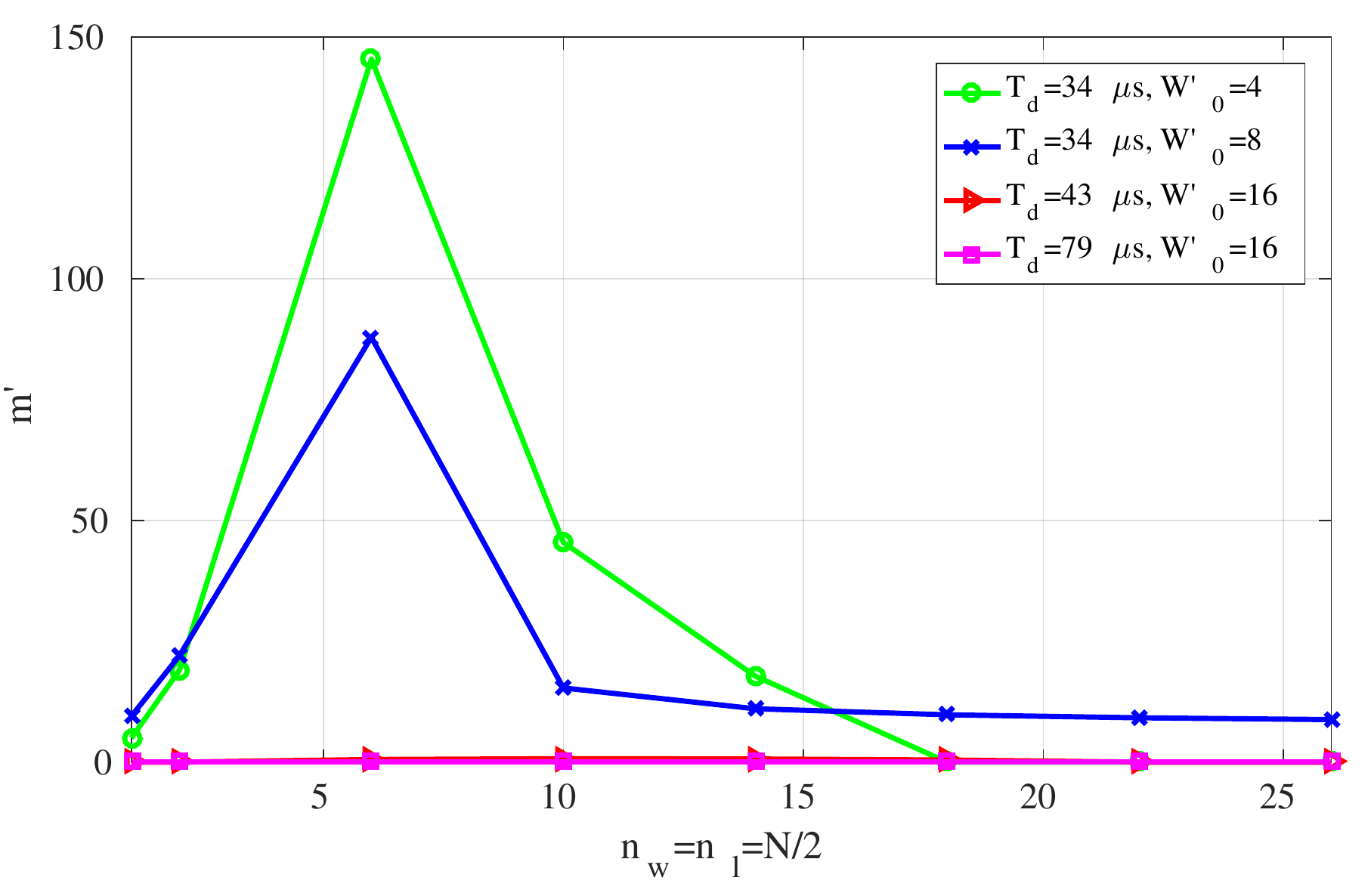}}
 \caption{The optimized $m'$ values for different number of nodes and different cases to achieve Wi-Fi access fairness.}
 \label{fig: sim0}
\end{figure}

Fig.~\ref{fig: sim0} shows the optimized LTE-LAA retransmission stage ($m'$) obtained by solving the eq (\ref{eq: Opt}) for different channel access parameters, for $e_l = 1$. When $W'_0 \le W_0$ and $T_d$ = DIFS, the optimized $m'$ must be very large to achieve access fairness. This is because the LTE-LAA with smaller $W'_0$ gets more frequent access to the channel and with larger $m'$ the probability of access decreases. However, for larger $W'_0$ (equal to Wi-Fi values) and $T_d$ (larger than DIFS), the LTE-LAA gets lower access to the channel and the retransmission ($m'$) should decrease to increase the access probability. In this way, the access probability of Wi-Fi in coexistence network is equal to Wi-Fi in Wi-Fi only network. 

\textbf{Summary:}
\begin{itemize}
    \item Access fairness may not be a good criterion because, for smaller $W'_0$, the optimized $m'$ for fairness is very large and for larger $W'_0$ and $T_d$, the optimal value is very small ($m'=0$). 
    
    \item Even if the access fairness is achieved, due to the airtime and data rate difference between the two systems, access fairness does not imply throughput fairness. 
 
\end{itemize}


\section{Proportional Throughput Fairness in Coexistence}

In the previous section, the LTE-LAA retransmission parameter ($m'$) and TXOP were optimized to achieve access (and throughput) fairness consistent with 3GPP definition, without any consideration of the impact on LTE-LAA throughput. 
Therefore, using the notion of proportional fairness - which considers the performance of {\em both} networks - may provide a more appropriate definition of fairness, and is investigated in this section. In a WLAN network in which the APs and UEs follow the CSMA/CA protocol but may use different data rates, proportional fairness is achieved by tuning the minimum contention window size or LTE-LAA TXOP to assign equal {\em airtime} (and not equal access) to different nodes \cite{WLAN_Fairness2005, Fairness_11e}.

Consistent with the above, proportional fairness for coexistence suggests that LTE-LAA parameters can be tuned to allocate bandwidth to each of the networks (Wi-Fi and LTE-LAA) in consideration of their (different) data rates and channel access mechanisms. Although the channel access mechanism is qualitatively the same for LTE-LAA and Wi-Fi, the respective key parameters (notably TXOP and data rates) are different. So, in order to achieve fairness in such a heterogeneous network, the TXOP of LTE-LAA is a suitable parameter to be tuned to achieve the proportional fairness. We explore this next, using a system model similar to \label{3GPPfair} which is illustrated in Fig.~\ref{fig: Diag} (b). 

As explained in \cite{WLAN_Fairness2005}, proportional throughput fairness is achieved via 
\begin{equation}
\begin{split}
& \underset{T_D}{\max} \sum_{i\in \{w,l\}} \log(Tput_i(\alpha))=\underset{T_D}{\max}\left(\prod_{i\in \{w,l\}}Tput_i(T_D) \right),\\
& s.t.~~0 < T_D < 6~\text{ms},
\end{split}
\label{eq: Opt3}
\end{equation}
where the maximum $T_D=6$ ms meets the maximum TXOP constraint in Table~\ref{table: LAAclass1}. This can be solved by considering Wi-Fi and LTE-LAA throughput in eq (\ref{eq: tputw}) and (\ref{eq: tputl}).

\subsection{Numerical Results for Proportional Fairness}
\label{sec: NumPropFair}

\begin{figure}[t]
\setlength{\belowcaptionskip}{-0.1in}
\centerline{\includegraphics[width=3.3in]{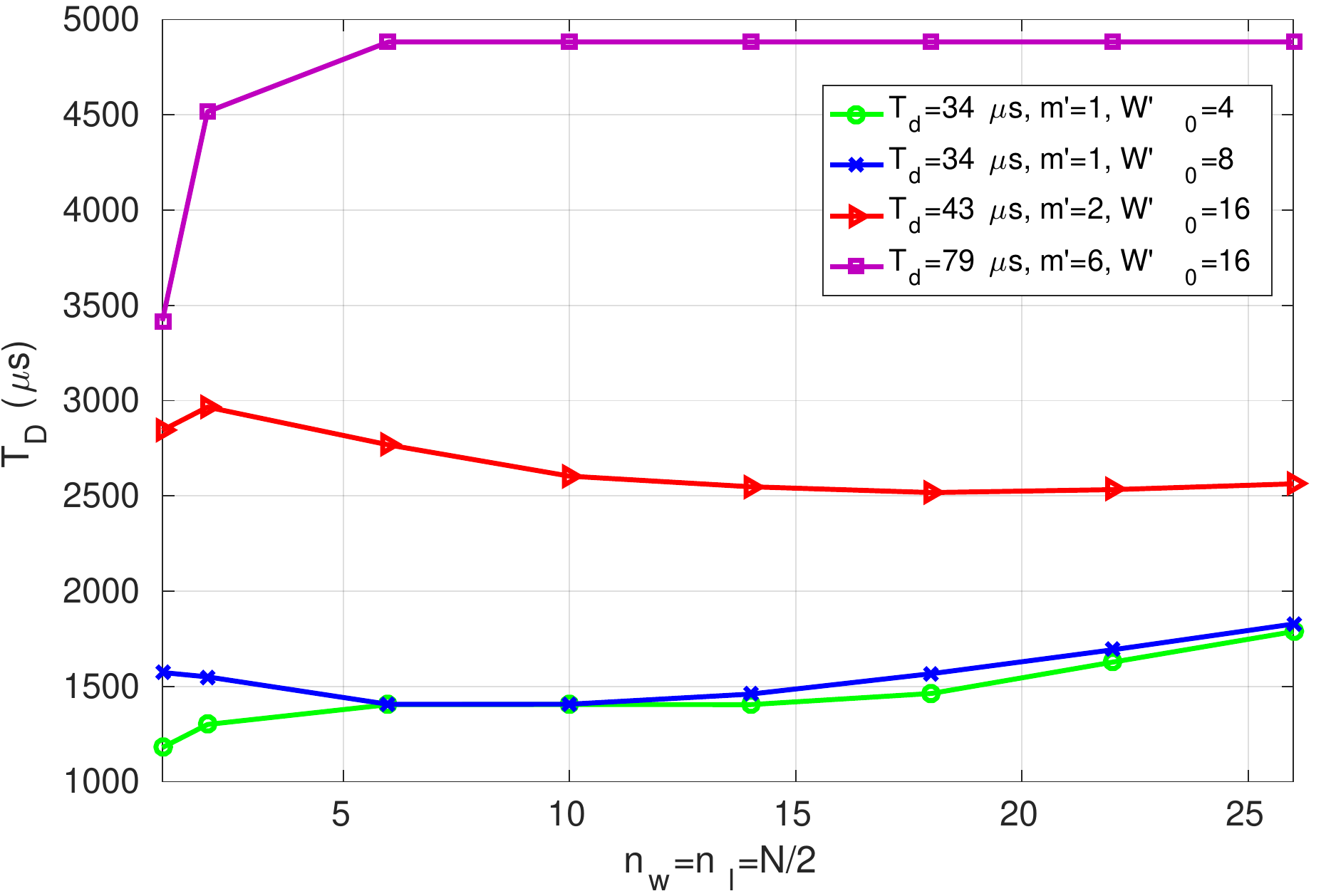}}
 \caption{The optimized LTE-LAA TXOP to achieve the proportional throughput fairness for the 9 Mbps Wi-Fi data rate and 7.8 Mbps LTE-LAA data rate. The Wi-Fi DCF parameters are fixed as: DIFS $=34\mu s$, $m=6$, and $W_0=16$.}
 \label{fig: simprop1}
\end{figure}

\begin{figure}[t]
\setlength{\belowcaptionskip}{-0.1in}
\centerline{\includegraphics[width=3.3in]{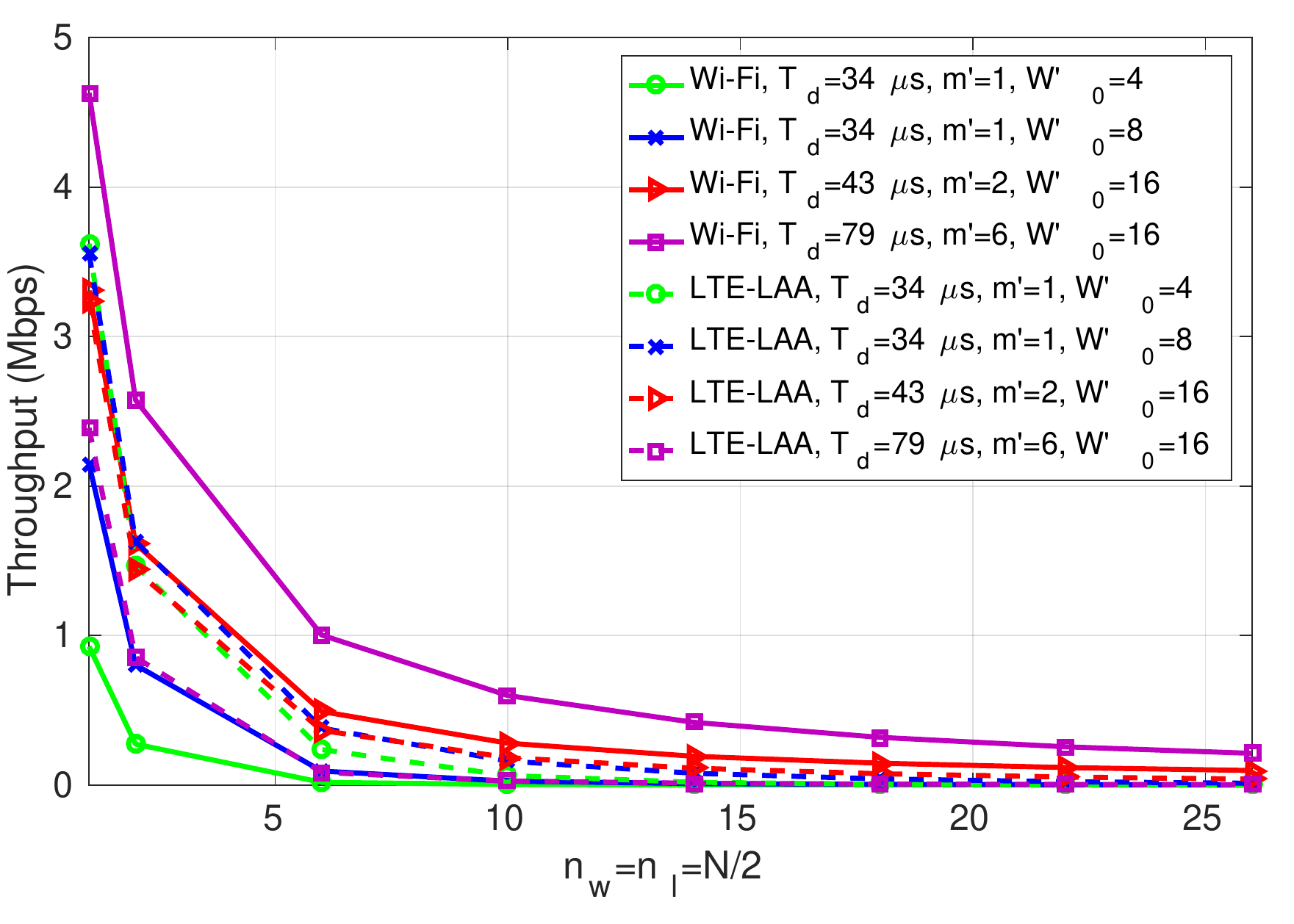}}
 \caption{The per-user throughput of Wi-Fi and LTE-LAA for the 9 Mbps Wi-Fi data rate and 7.8 Mbps LTE-LAA data rate when the proportional throughput fairness is achieved. The Wi-Fi DCF parameters are fixed as: DIFS $=34\mu s$, $m=6$, and $W_0=16$.}
 \label{fig: simprop2}
\end{figure}

Results for proportional fair coexistence are derived for the same scenario as in Section \ref{sec: Num3GPPFair}; the Wi-Fi parameters are illustrated in Table~\ref{table: WiFipar}. 
The data rate of Wi-Fi is 9 Mbps (airtime of $1.82$ ms) and LTE-LAA is 7.8 Mbps. For achieving the proportional fairness the TXOP of LTE-LAA is optimized by solving Eq. (\ref{eq: Opt3}). Fig.~\ref{fig: simprop1} illustrates the optimized TXOP of LTE-LAA. As the $T_d$ and LTE-LAA channel access parameters are changed (priority classes from 1 to 4), the optimized TXOP of LTE-LAA for proportional fair increases. This is expected because increasing the priority class implies decreasing channel access, thus the LTE-LAA TXOP should be larger to compensate for achieving proportional fairness.
Fig.~\ref{fig: simprop2} illustrates the achieved per user throughput of Wi-Fi and LTE-LAA when the proportional fairness is achieved. When the LTE-LAA transmits with higher priority classes (with larger $T_d$ and LTE-LAA channel access parameters), the throughput of Wi-Fi is larger than the throughput of LTE-LAA. Because Wi-Fi channel access parameters are smaller than LTE-LAA, Wi-Fi gets more access to the channel than LTE-LAA.

\begin{figure}[t]
\setlength{\belowcaptionskip}{-0.1in}
\centerline{\includegraphics[width=3.3in]{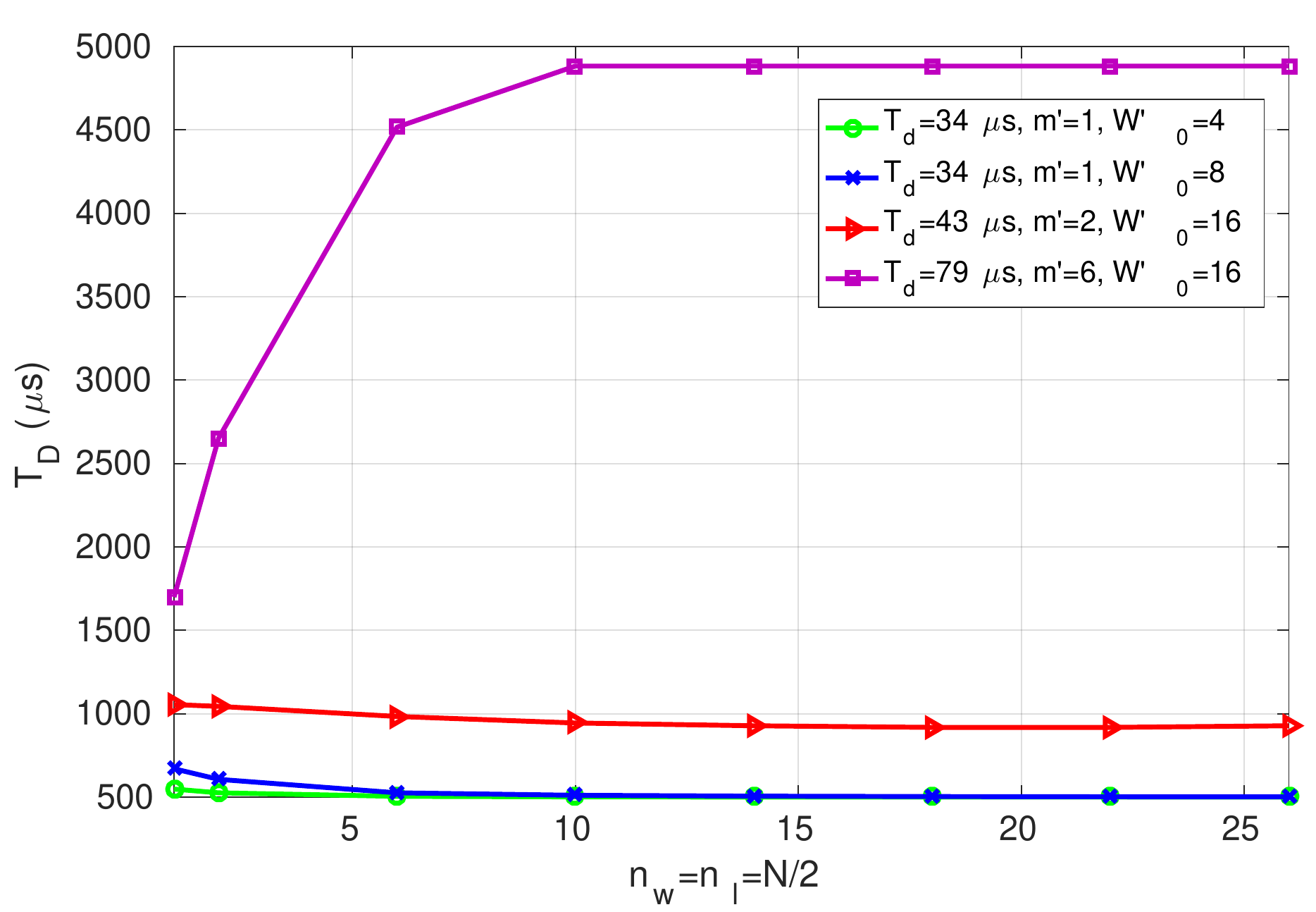}}
 \caption{The optimized LTE-LAA TXOP to achieve the proportional fairness for the 54 Mbps Wi-Fi data rate and 70.2 Mbps LTE-LAA data rate. The Wi-Fi DCF parameters are fixed as: DIFS $=34\mu s$, $m=6$, and $W_0=16$.}
 \label{fig: simprop3}
\end{figure}

\begin{figure}[t]
\setlength{\belowcaptionskip}{-0.1in}
\centerline{\includegraphics[width=3.3in]{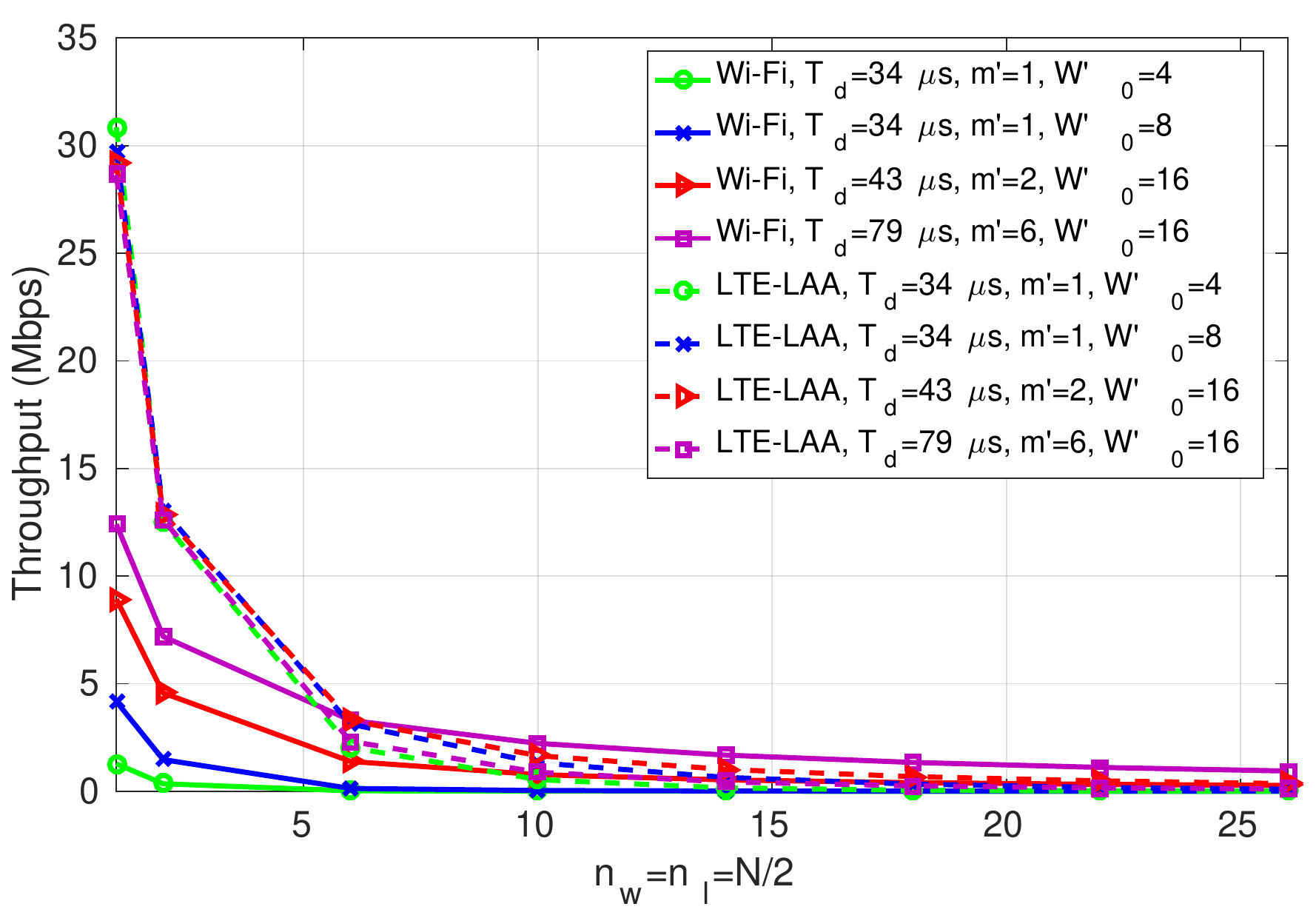}}
 \caption{The per-use throughput of Wi-Fi and LTE-LAA when the proportional fairness is achieved for the 54 Mbps Wi-Fi data rate and 70.2 Mbps LTE-LAA data rate. The Wi-Fi DCF parameters are fixed as: DIFS $=34\mu s$, $m=6$, and $W_0=16$.}
 \label{fig: simprop4}
\end{figure}

Fig.~\ref{fig: simprop3} and Fig.~\ref{fig: simprop4} illustrates the optimized LTE-LAA TXOP and achieved per user throughput based on the proportional fairness for Wi-Fi data rate of 54 Mbps and LTE-LAA data rate of 70.2 Mbps, respectively. Similar observation as the previous scenario with lower data rate holds with the difference that for the LTE-LAA priority class 1 to 3, the LTE-LAA TXOP is much smaller; the higher data rate of Wi-Fi in comparison with Fig.~\ref{fig: simprop2} makes the Wi-Fi frame airtime very small and decreases  Wi-Fi throughput. The Wi-Fi throughput increases when coexisting with the higher LTE-LAA priority classes indicate that under proportional fairness, the LTE-LAA with higher priority classes achieves greater throughput fairness with Wi-Fi.

\textbf{Summary:}
\begin{itemize}
    \item In a proportional fair regime (in contrast to the 3GPP notion of fairness), the throughput of Wi-Fi and LTE-LAA are not zero in any of the LTE-LAA priority classes. Depending on the LTE-LAA priority class, the Wi-Fi throughput is larger or smaller than the LTE-LAA. This fact implies that the proportional fairness is a better notion of fairness than 3GPP fairness in which the LTE-LAA throughput performance of priority class 1 and 2 was zero.  
    
    \item Generally, the optimized TXOP of the LTE-LAA for the achieved proportional fairness does not follow the suggested TXOP defined in \cite{ETSILAA17} as illustrated in Table \ref{table: LAAclass1}. But, priority class 1 and 4 are closer to the parameters of the Table.
\end{itemize}

\section{Wi-Fi TXOP: Effect on Coexistence Fairness}

Wi-Fi systems at higher data rates (e.g. 802.11ac) are allowed larger airtime per access, due to aggregation of MAC Protocol Data Unit (A-MPDU) frames that are allowed to transmit for a Wi-Fi TXOP duration. The maximum MPDU size is 11454 bytes with 4 bytes of MPDU delimiter and the maximum Wi-Fi TXOP for an A-MPDU frame is 5.46 ms in the very high throughput (VHT) mode \cite{11acAMPDU1, 11acAMPDU2}. Similar to the Fig.~\ref{fig: WiFitime}, the MAC header for each MPDU is 30 bytes and FCS is 4 bytes at the end of each MPDU which is part of the MPDU size. The VHT frame preamble length is 10 OFDM symbols. The acknowledgment of the A-MPDU frame is sent through the block ACK request (BAR) from the transmitter node and block ACK (BA) from the receiver node. We assume that because of synchronized collision if a collision happens, the total A-MPDU would be in error and should be retransmitted. The throughput calculation in Section \ref{sec: TputCal} is thus valid except that eq. (\ref{eq: TsTc}) for the average duration of a successful and collision events should be updated as follows (to capture the effect of transmission for Wi-Fi TXOP duration and BAR/BA transmission):
\begin{equation}
\begin{split}
T_{sw}&=\text{PhyH}+\text{MACH}+\text{Psize}+\text{SIFS}+\text{BAR}+\text{SIFS}\\&+\text{BA}+\text{DIFS},\\
T_{cw}&=T_{sw},
\end{split}
\label{eq: TsTc2}
\end{equation}

The updated parameters of Wi-Fi in VHT mode is represented in Table~\ref{table: WiFi11acpar}.
We consider a scenario with the Wi-Fi data rate of $r_w=78$ Mbps (MCS8 with 256-QAM and code rate of 3/4) and A-MPDU sizes of $N_{MPDU}=2$ and 4 where their corresponding Wi-Fi TXOP are 2.39 ms and 4.74 ms, respectively. 

\begin{table}
\caption{Wi-Fi VHT parameters.}
\label{table: WiFi11acpar}
\begin{center}
\begin{tabular}{|c|c|}
\hline 
Parameter & value \\
\hline
\hline
 $r_0$ & 26 Mbps \\
\hline
 PhyH & 40 $\mu s$ \\
 \hline
 MACH & (38 bytes)$/r_w$ $\mu s$\\
 \hline
 BAR & 20+(24 bytes)$/r_0$ $\mu s$ \\
  \hline
 BA & 20+(32 bytes)$/r_0$ $\mu s$ \\
   \hline
 $N_{MPDU}$ & 2 or 4 \\
  \hline
 $N_B$ & $N_{MPDU}\times 11416$ (bytes) \\
  \hline
 Psize & $N_B/r_w$ $\mu s$\\
 \hline
\end{tabular}
\end{center}
\end{table}

\begin{figure}[t]
\setlength{\belowcaptionskip}{-0.1in}
\centerline{\includegraphics[width=3.3in]{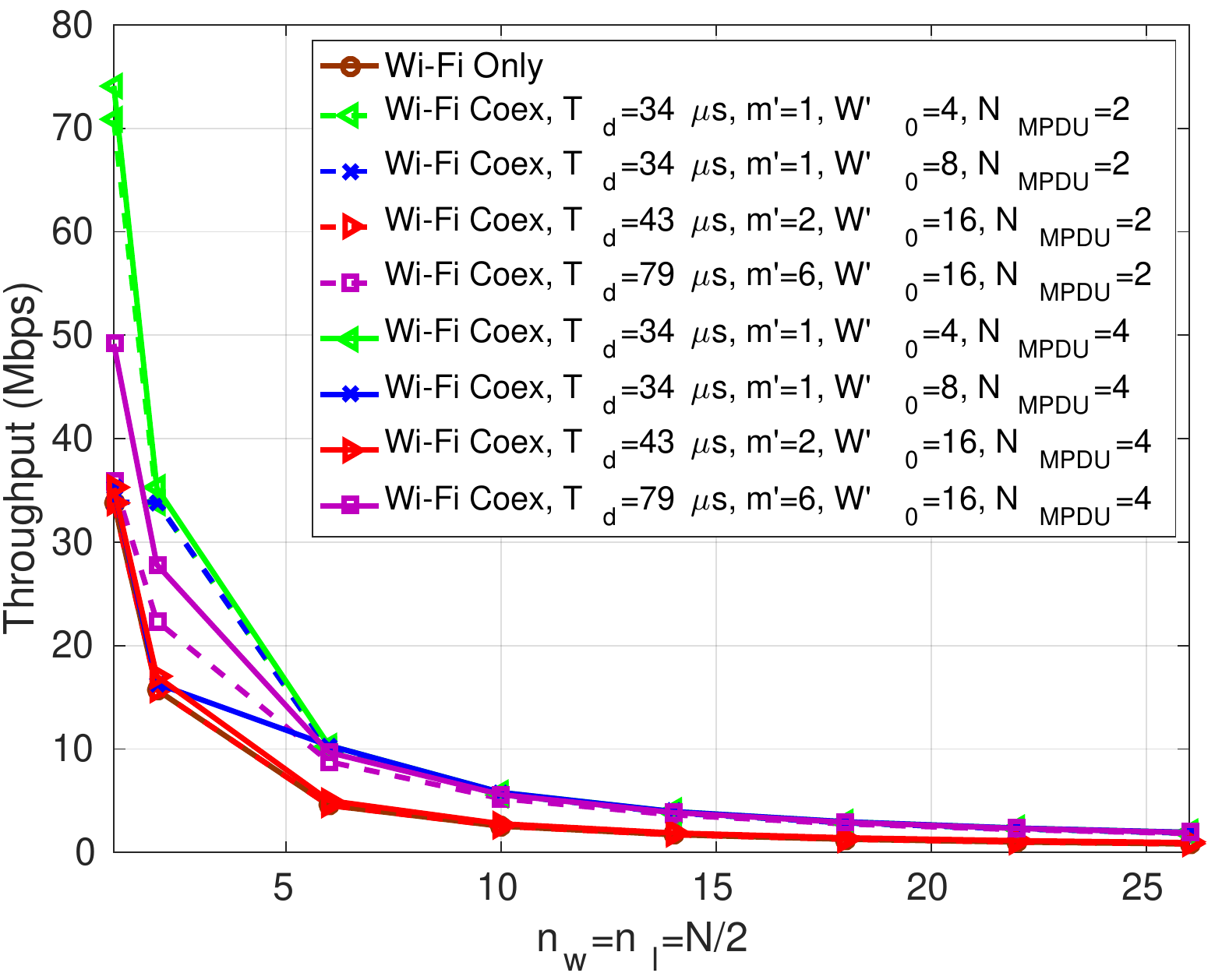}}
 \caption{The per-user throughput of Wi-Fi in VHT mode under the 3GPP notion fairness with the 78 Mbps VHT Wi-Fi data rate and 70.2 Mbps LTE-LAA data rate. The Wi-Fi DCF parameters are fixed as: DIFS $=34\mu s$, $m=6$, and $W_0=16$.}
 \label{fig: simtxop1}
\end{figure}

\begin{figure}[t]
\setlength{\belowcaptionskip}{-0.1in}
\centerline{\includegraphics[width=3.3in]{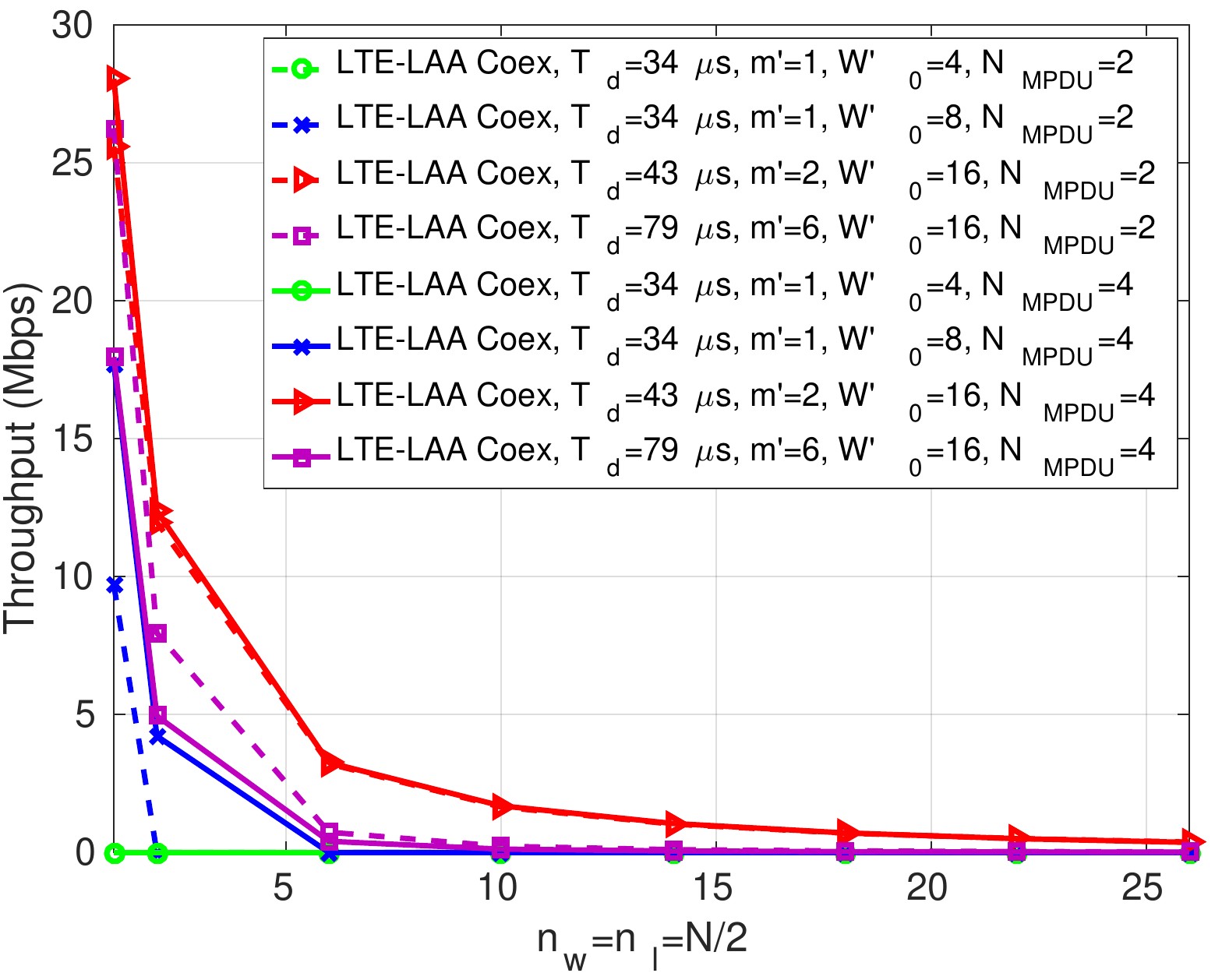}}
 \caption{The per-user throughput of LTE-LAA under the 3GPP fairness for the 78 Mbps VHT Wi-Fi data rate and 70.2 Mbps LTE-LAA data rate. The Wi-Fi DCF parameters are fixed as: DIFS $=34\mu s$, $m=6$, and $W_0=16$.}
 \label{fig: simtxop2}
\end{figure}

Fig.~\ref{fig: simtxop1} illustrates the analytical throughput of Wi-Fi and Fig.~\ref{fig: simtxop2} shows the analytical throughput of LTE-LAA under the 3GPP throughput fairness for $N_{MPDU}$ of 2 and 4. To achieve the 3GPP fairness, eq. (\ref{eq: Opt2}) is solved to find the optimize LTE-LAA TXOP. Similar to Section \ref{sec: Num3GPPFair}, the LTE-LAA throughput for priority class 1 is almost zero but for priority class 2 the LTE-LAA achieves a higher throughput for few number of nodes with both $N_{MPDU}$ values (i.e. the optimized LTE-LAA TXOP is larger); this implies that larger TXOP of Wi-Fi helps the LTE-LAA to achieve a higher throughput because it allows the LTE-LAA to transmit with higher airtime. For the priority class 3 and 4, the observation is the same as Section \ref{sec: Num3GPPFair}. Generally speaking, Wi-Fi nodes in coexistence achieve a higher throughput than LTE-LAA only under the 3GPP fairness definition; in these scenarios, LTE-LAA achieve very low throughput (LTE-LAA TXOP $ = 0$ as explained in Section \ref{sec: Num3GPPFair}), thus reducing contention to just between the Wi-Fi nodes which are half of the nodes in Wi-Fi only network (i.e., $n_w=N/2$).

\begin{figure}[t]
\setlength{\belowcaptionskip}{-0.1in}
\centerline{\includegraphics[width=3.3in]{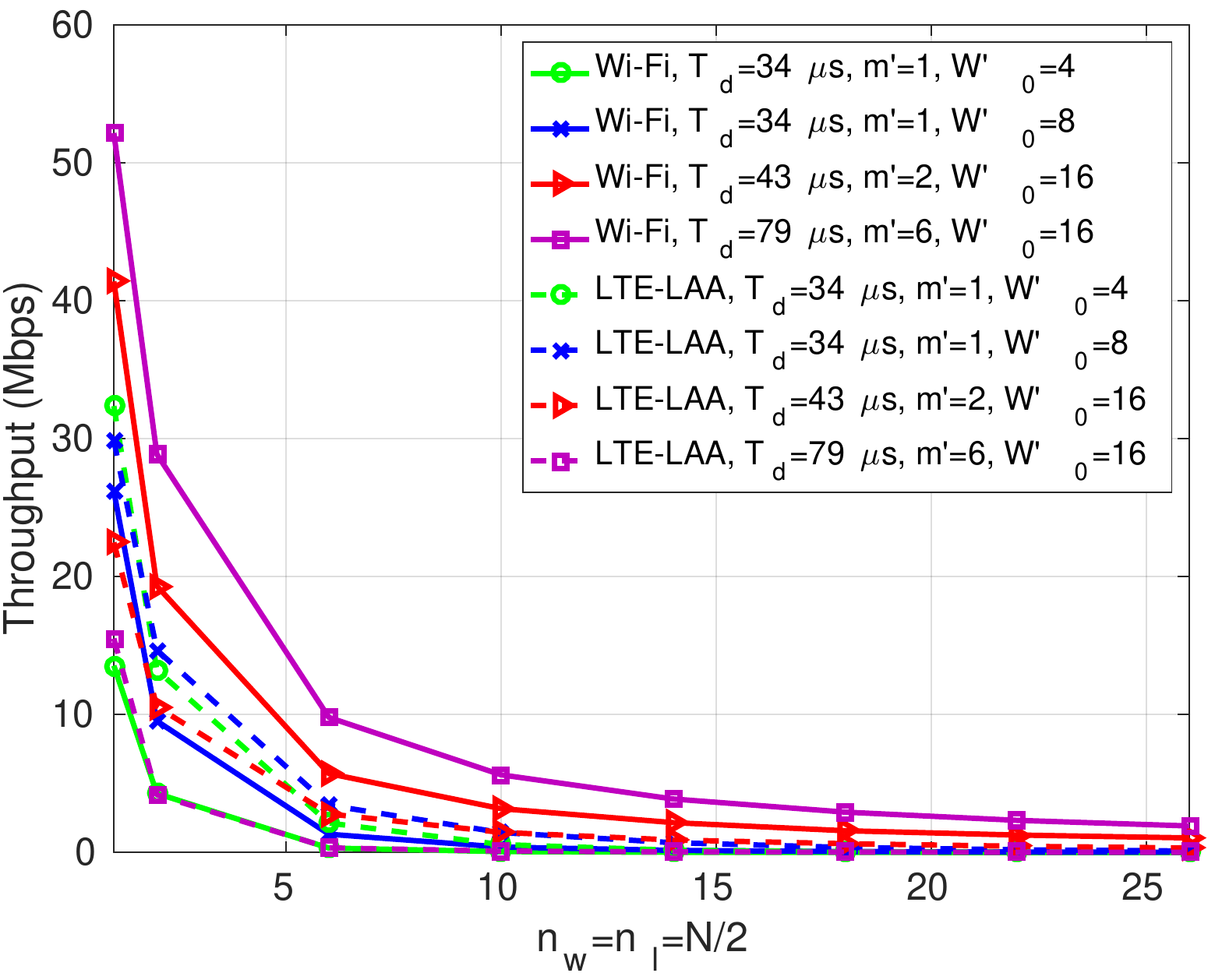}}
 \caption{The per-user throughput of Wi-Fi in VHT mode and LTE-LAA under the proportional fairness with the 78 Mbps VHT Wi-Fi data rate and 70.2 Mbps LTE-LAA data rate. The Wi-Fi DCF parameters are fixed as: DIFS $=34\mu s$, $m=6$, and $W_0=16$.}
 \label{fig: simtxop3}
\end{figure}

Fig.~\ref{fig: simtxop3} shows the throughput performance of Wi-Fi in VHT mode and LTE-LAA under proportional fairness. To achieve the proportional fairness, eq. (\ref{eq: Opt3}) is solved to find the optimized LTE-LAA TXOP. Similar observation as Section \ref{sec: NumPropFair} can be made except that when Wi-Fi has a higher data rate, the Wi-Fi TXOP remains the same (Wi-Fi frame airtime remains the same), thus the Wi-Fi throughput is not small and depending on the LTE-LAA priority classes (contention parameters) Wi-Fi could achieve a comparable throughput. Wi-Fi achieves a higher throughput when contending with LTE-LAA at priority class 3 and 4 and lower throughput when contending with LTE-LAA at priority class 1 and 2. 

\textbf{Summary:}
\begin{itemize}
    \item The Wi-Fi nodes in VHT mode can transmit larger Wi-Fi frames for the duration of TXOP. This helps the coexistence system to achieve a better throughput fairness in both 3GPP and proportional notion of fairness. 
    
    \item When Wi-Fi transmits for TXOP duration, the LTE-LAA in some of the priority classes suffers from the zero throughput using the 3GPP fairness. However, the proportional fairness illustrates a better throughput fairness and trade-off of throughput between Wi-Fi and LTE-LAA at different priority classes of LTE-LAA.
    
\end{itemize}

\section{Conclusion}

In this work, a modified analytical framework for estimating the throughput of Wi-Fi/LTE-LAA coexistence which accounts for different parameters of the two networks is developed. Using this, the throughput fairness based on the 3GPP definition is studied. The 3GPP fairness results are meaningfully achievable only for LTE-LAA with higher priority classes. Consequently, we moved to the exploration of other well-respected notions of fairness - notably access and proportional fairness as applied to coexistence. 
The results conclusively show that proportional fairness is a much better notion than 3GPP fairness and produces equitable results for both networks in a larger variety of scenarios.

\section*{ACKNOWLEDGMENT}
This work was supported by the National Science Foundation (NSF) under grant 1618920.

\bibliographystyle{IEEEtran}
\bibliography{WiFiCoex}

\end{document}